# Strain effects on oxygen vacancy formation energy in perovskites

Tam Mayeshiba[1]
Dane Morgan[1*]

[1]Department of Materials Science and Engineering, University of Wisconsin-Madison, 1509 University Ave, Madison, WI, 53706
* Corresponding author

## Abstract

Oxygen vacancy formation energy is an important quantity for enabling fast oxygen diffusion and oxygen catalysis in technologies like solid oxide fuel cells. Both previous literature in various systems and our calculations in $LaMnO_3$, $La_{0.75}Sr_{0.25}MnO_3$, $LaFeO_3$, and $La_{0.75}Sr_{0.25}FeO_3$, show mixed results for the direction and magnitude of the change in vacancy formation energy with strain. This paper develops a model to make sense of the different trend shapes in vacancy formation energy versus strain. We model strain effects using a set of consistent ab initio calculations, and demonstrate that our calculated results may be simply explained in terms of vacancy formation volume and changes in elastic constants between the bulk and defected states. A positive vacancy formation volume contributes to decreased vacancy formation energy under tensile strain, and an increase in elastic constants contributes to increases in vacancy formation energy with compressive and tensile strains, and vice versa. The vacancy formation volume dominates the linear portion of the vacancy formation energy strain response, while its curvature is governed by the vacancy-induced change in elastic constants. We show results sensitive to B-site cation, A-site doping, tilt system, and vacancy placement, which contributions may be averaged under thermally averaged conditions. In general, vacancy formation energies for most systems calculated here decreased with tensile strain, with about a 30-100 meV/% strain decrease with biaxial strain for those systems which showed a decrease in vacancy formation energy. Experimental verification is necessary to confirm the model outside of calculation.
**Keywords:** perovskite; vacancy formation energy; strain; SOFC

## 1. Introduction

Fast oxygen conduction is important in devices that need to move oxygen rapidly, like solid oxide fuel cells (SOFCs), with perovskites being one of the most widely used classes of fast oxygen-ion conductors.[1, 2] The most common method of obtaining fast oxygen transport in perovskites is through optimizing the chemistry, but strain is another possible approach, as has been reviewed in Ref. [3]. As oxygen diffusion in perovskites is generally mediated by oxygen vacancies, the primary contributors to fast oxygen diffusion are a high concentration of mobile (generally disordered) oxygen vacancies, and a low effective oxygen migration barrier. Our previous calculations showed that tensile strain lowers oxygen migration barriers in perovskites.[4, 5] The effects of strain on oxygen vacancy concentration are less clear.

Aliovalent doping is the primary method for forming mobile oxygen vacancies for fast oxygen-ion conducting perovskites in major applications.[1, 2] However, the concentration of vacancies can change dramatically with temperature and oxygen partial pressure, for example, as shown in Ref. [6]. Therefore, low oxygen vacancy formation energy can still be a critical property to enable fast oxygen diffusion, particularly at lower operating temperatures. Oxygen vacancy content is also important in its own right, for example surface oxygen vacancy content for catalytic activity.[7] However, literature gives conflicting reports on the extent, direction, and mechanism of the strain effect on oxygen vacancy content and oxygen vacancy formation energy.

Computational studies of oxygen vacancies in perovskites with biaxial strain typically focus on the the two symmetry inequivalent vacancy sites in the cubic structure (space group #221, *Pm-3m*[8]), which sites are generally referred to as in-plane (IP) and out-of-plane (OP), shown in Fig. 1. The studies of in-plane (IP)[9-12] and out-of-plane (OP)[9, 10, 12, 13] vacancies have found increasing tensile strain to increase,[11] decrease,[7, 9, 11-13] or sometimes have no apparent effect[9, 11] on oxygen vacancy formation energy. Sometimes the effect is near linear,[7, 10, 11] and sometimes it is curved.[9, 11, 13]

Experimentally, there is only one study on perovskites of which we are aware that directly determined the strain response of vacancy concentration. Specifically, Petrie *et al.* found biaxial tensile strain to increase oxygen non-stoichiometry (that is, increasing $\delta$ in $SrCoO_{3-\delta}$), also corresponding to a computationally calculated 30% decrease in oxygen activation barrier (the migration part) at 2% tensile strain, which may aid in oxygen loss as oxygen migrates more easily out of the perovskite layer.[14]

Studies in the defected-fluorite structure, which like the perovskite structure is also a fast oxygen-ion conducting structure, have generally shown biaxial tensile strain to decrease oxygen vacancy formation energy. A recent study by Balaji Gopal et al. in $CeO_2$ shows biaxial strain producing a non-monotonic decrease in oxygen vacancy formation energy in both ab-initio computation and in interpretation of experiment.[15] Sheth et al. applied strain using temperature changes and found a decrease in enthalpy of reduction with tensile strain, which they relate to a decrease in defect formation energy, as defect formation energetics move from a bulk model to a thin film model.[16] Other computational studies of ceria show oxygen vacancy formation energy decreasing linearly as biaxial strain progresses from compression to tension.[17, 18]

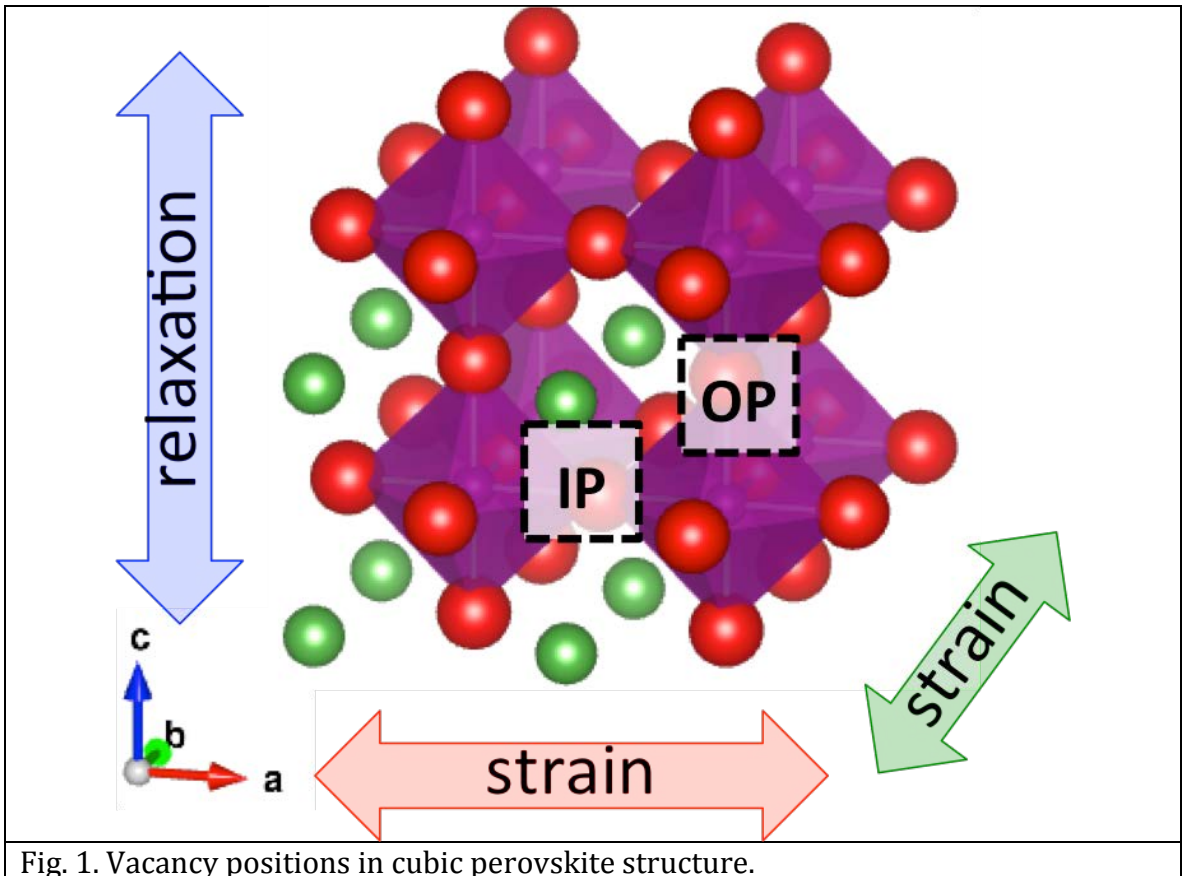

Fig. 1. Vacancy positions in cubic perovskite structure.

There are various theories to explain the observed vacancy formation energy responses to strain, although no single theory appears to be able to explain all the responses. The major categories of explanations, with some degree of interrelation and interdependence, are (1) volume, where positive oxygen vacancy formation or vacancy relaxation volume (taken as a total contribution from vacancy-creation contraction and cation-reduction expansion)[19, 20] reacts differently to tensile and compressive strains and to IP or OP vacancy placement;[9] (2) bonding, where bonding and antibonding orbitals change in energy with strain, and energy contributions are lost and/or gained,[9, 14] or where stronger bonding as observed from charge density maps increases vacancy formation energy with strain;[10] (3) internal relaxation of the perovskite, where octahedral tilting absorbs strain,[9] or where distortions compete,[7] or where relaxation effects in different systems produce different behaviors;[11] and (4) elastic constants, where defects weaken the elastic modulus, so defected systems can be strained more easily, or strained films form vacancies more easily, in perovskites[13] or defected fluorites.[21]

This work shows that one can characterize the vacancy formation energy as a function of strain, $E_{vf}$(strain), in terms of a few intuitive parameters related to vacancy volume and changes in elastic constants associated with creating the vacancy. Such a simple model enables researchers to understand why $E_{vf}$(strain) for different materials can show a variety of behaviors, including increases or decreases and quasi-linear or parabolic curves, in terms of simple physical parameters. This model allows researchers to fit $E_{vf}$(strain) with just a few parameters. By reducing $E_{vf}$(strain) to a few parameters, researchers can more easily use intuition, machine learning, and other tools to predict and screen for new materials with desired vacancy energetics.

## 2. Material and methods

### 2.1. Calculation details

All ground-state energy calculations were performed using the Vienna Ab-initio Simulation Package (VASP)[22-25] with workflows managed through the MAterials Simulation Toolkit (MAST).[26] PAW-PW91[27-31] pseudopotentials were chosen as described in Ref.[4]. A 2x2x2 (40-atom) LaXO$_3$ starting supercell was used unless otherwise noted. For doped supercells, Strontium doping took place on body-diagonal A-site positions 1 and 8 (see Fig. S1 of Ref. [4] for positions). Notation is LXO for undoped supercells LaXO$_3$ and LSX for doped supercells La$_{0.75}$Sr$_{0.25}$XO$_3$, where X is the first letter of the B-site cation's atomic symbol. Calculations are done with the GGA+U method,[29, 32, 33] with *U-J* values given in Table 1, and a 6x6x6 Monkhorst-Pack[34] kpoint mesh. Energies for an undefected supercell are converged to within about 1 meV/atom with respect to an 8x8x8 kpoint mesh at 0% strain, and to within about 10 meV/atom at -2% to +3% biaxial strain.

Although the materials studied here are expected to be paramagnetic (PM) at high temperatures,[35] e.g. the operating temperatures of solid oxide fuel cells, the materials are modeled as ferromagnetic (FM). The FM model is much simpler than a paramagnetic model, which makes the calculations tractable, and has been argued to give similar energetics.[36] For unstrained materials, Lee et al. found oxygen vacancy formation energies for FM structures, AFM structures, and high-temperature experiment to be similar to each other, with FM vacancy formation energies within the experimental error bars of about 0.7 eV for LaMnO$_3$ and LaFeO$_3$ with GGA+U (Ref. [36] Figure 11b). The effects of strain on an AFM or PM structure may still result in different curves than those shown here. However, even for the strong AFM LaFeO$_3$ the difference between FM and AFM vacancy energy strain response is relatively minor (see the Supplementary Information (SI) Section S3), suggesting that the choice of FM state has a weak effect on the results of this paper. Furthermore, we expect the principles of the model derived in this work to remain the same regardless of the magnetic state.

This study fixes the pseudo-cubic 2x2x2 unit cell of 40 atoms to be a starting cubic unit cell with cubic lattice vectors, with lattice parameters *a*=*b*=*c* and at 90 degrees, and then allowing relaxation in lattice parameter *c* along the orthogonal *z* direction. This approach mimics the pseudomorphic relaxation of an epitaxially strained layer on a cubic substrate, e.g. on $SrTiO_3$[8] or on a cubic fluorite. This approach has been used frequently in previous studies.[9, 10, 12, 37-49]

The initial pseudo-cubic structure was determined as follows. First, fractional coordinates were generated by the SPUDS program[50] with the Jahn-Teller distortion setting on. B-site cations Mn and Fe were started with the $a^-b^+a^-$ tilt system (spacegroup #62, *Pbnm*),[8, 51] while B-site cation Ni (appendix only) was started with the $a^-a^-a^-$ tilt system (spacegroup #167, *R-3c*).[8] The resulting supercells were orthorhombic and trigonal, respectively. These supercells were then converted into a pseudo-cubic supercell using the algorithm in Ref. [52] (where *U* and *U'* are column vectors) and transformation methods in the pymatgen library;[53] see SI Section S5 for details. Then, the pseudo-cubic lattice parameter was imposed for the length of all three orthogonal lattice parameters. Finally, VASP was used for a sequential set of volume-only, internal, volume-only, and internal relaxations, generating a relaxed pseudo-cubic supercell. Table 1 shows the obtained lattice parameters.

In the following text, "defect" refers only to an oxygen vacancy defect, and the cell referred to as "undefected" can be doped or undoped. For 40-atom cells, a single oxygen defect is created, giving an oxygen vacancy concentration of 1/24 ($\delta$=0.125 for perovskite formula $ABO_{3-\delta}$). The defect properties determined in this work are therefore most accurate near the 1/24 concentration range, and cannot necessarily be used at other defect concentrations. For example, SI Section S2 shows significantly different defect energies when using a 4x4x4 cell, which has a defect concentration of 1/192 ($\delta$=0.015625).

At each strain state, including zero strain, strain relative to the pseudo-cubic lattice parameter in Table 1 was applied to the undefected supercell along orthogonal lattice parameters *a* and *b* with an opposite sign strain of the same magnitude for lattice parameter *c* as a starting guess (e.g. for 1% tensile strain, 1% compressive strain along c was set as a starting value). A custom z-relaxation-only compilation of VASP was used to relax orthogonal lattice parameter *c* along with all internal atomic coordinates.[54] No atomic movement was restricted, although not all distortions (e.g. small A-site cation distortions) may be physical.[47] Using these new fixed lattice parameters at each strain state, IP and OP defects were created at oxygen positions 30 (with a bond in the plane of the strain) and 29 (an apical oxygen), respectively (see Fig. S1 of Ref. [4] for positions), and both those defected states and the undefected state were allowed to relax internally again. Finally, energies were calculated with a static calculation.

In the internally relaxed supercells, there are actually several symmetry-inequivalent vacancy positions in each of the IP and OP categories. However, the differences between the two categories are expected to be larger and more informative than variations within each category, so only a single representative IP vacancy and a single representative OP vacancy are considered here.

Since B-site cations Mn and Fe have directionally dependent tilt systems, both orientations of the tilt system with respect to the plane of strain were evaluated, that is, a$^-$b$^+$a$^-$ or a$^-$a$^-$b$^+$ with respect to the x, y, and z axes. The designation "bulk tilt" in plots signifies the tilt system with the lowest calculated bulk energy at each strain. Tilt systems in the text will be abbreviated as *aba*, *aab*, and *aaa* for a$^-$b$^+$a$^-$, a$^-$a$^-$b$^+$, and a$^-$a$^-$a$^-$, respectively, as *a* is always a$^-$ and *b* is always b$^+$.

Table 1. Cubic lattice parameters for starting structures and *U-J* value

| System | Calculated cubic-constrained unit cell lattice parameter (Å) | *U-J* value for B-site cation[55] |
|---|---|---|
| LFO | 3.976 | 4.0 |
| LSF | 3.951 | 4.0 |
| LMO | 3.967 | 4.0 |
| LSM | 3.939 | 4.0 |
| LNO | 3.881 | 6.4 |
| LSN | 3.872 | 6.4 |

## 2.2. Vacancy formation energy

Vacancy formation energy is calculated through the method in Lee et al,[36] which in this particular case amounts to Eq. 1, where *E* refers to the supercell energy, $\mu_O$ has a constant value of -5.86 eV for the soft oxygen PAW-PW91 pseudopotential at 1173K and 0.1 atm *pO$_2$*, subscript *vf* indicates vacancy formation, subscript *d* indicates the defected supercell, and subscript *u* indicates the undefected supercell.

| $E_{vf} = E_d - E_u + \mu_O$ | Eq. 1 |
|---|---|

We note that these formation energies are in fact equal to formation enthalpies, assuming that the external pressure impacting the c-axis relaxation is negligible. However, they will simply be referred to as formation energies throughout this paper. To accelerate the calculations and reduce instability we keep the c-axis fixed at the undefected cell value when we introduce the defect (a and b axes are also fixed due to the biaxial strain condition). Defects are all calculated within a neutral unit cell and are therefore not what are often called "charged" defects in ab initio models. This approach is appropriate for metallic or near-metallic systems like LMO, LSM, and LSF, none of which had a bandgap in our calculations. It is possible that LFO, which has a larger band gap (calculated as 0.35 eV in this work, but ranges up to 1.36 eV in other works, depending on exact supercell and calculation parameters[36, 56, 57]) would form charged defect states, depending on the Fermi level. However, we have not included these additional states to reduce the complexity of the calculations and to enable direct comparison across all the systems.

## 2.3. Vacancy formation energy as the difference between two parabolas

Considering vacancy formation energy as the difference between two parabolas simplifies the understanding of the varying effects of strain on vacancy formation energy. When both the defected and undefected energy versus strain curves are reasonably well modeled by parabolas, as will be shown in the Results section, each energy curve can be modeled as Eq. 2, where $\epsilon$ is strain, $a$ is the curvature, $h$ is the horizontal vertex offset (in units of strain), and $k$ is the vertical vertex offset (relating to overall stability).

| | |
|---|---|
| $E_{VASP}(\epsilon) = a(\epsilon - h)^2 + k$ | Eq. 2 |

Calculating vacancy formation energy using Eq. 1 with constituent energies of the parabolic form in Eq. 2 produces a parabola given by Eq. 3, with $\Delta a$ defined in Eq. 4. Appendix A contains more details and additional clarification.

| | |
|---|---|
| $E_{vf}(\epsilon) = \Delta a(\epsilon - h_{vf})^2 + k_{vf}$ | Eq. 3 |
| $\Delta a = a_d - a_u.$ | Eq. 4 |

The units of $\Delta a$ and $h_{vf}$ are eV/strain$^2$ and strain, respectively. However, it should be remembered that their values are dependent on the changes in defect concentration. For all values in the paper this change is 1/24, and SI Section S2 shows values for a change in concentration of 1/192. While it is temping to give the values normalized per unit change of concentration, there is insufficient study at this point to assess how they depend on concentration, so we simply give the values obtained for our specific concentrations here. This issue is discussed further in SI Section S2. Any quantity in this paper related to the changes with formation of the defects should be understood in the same way as just discussed, i.e., as the changes associated with creating a defect concentration of 1/24.

### 2.4. Relationship between $\Delta a$ and change in bulk modulus, $\Delta K$

The parabolic coefficient $\Delta a$ is proportional to a change in elastic behavior (see Appendix B for a derivation). Elastic constants may be expected to weaken with the introduction of a vacancy,[13] as has been shown in fluorite-type films,[58] leading to $\Delta a$ smaller than zero for the majority of systems.

Assuming an isotropic elastic material, $a$ relates directly to the bulk modulus $K$ and Poisson's ratio $\nu$ through Eq. 5, where $V_0$ is the unstrained supercell volume. See Appendix B for relationships to other elastic constants.

| | |
|---|---|
| $a = \dfrac{3V_0 K(1 - 2\nu)}{(1 - \nu)}$ | Eq. 5 |

Assuming little change in Poisson's ratio from the bulk to defected state (which is the approximation made in the derivation), $\Delta a$ therefore represents a change in bulk modulus between the defected and undefected state, as in Eq. 6. These changes in bulk modulus can be quite significant and can be estimated from Eq. 6. A typical approximate supercell volume of 500Å³, Poisson's ratio of about 0.33, and $\Delta a$ of 100 eV/strain² corresponds to a change in bulk modulus of approximately 20 GPa, which is a significant and measurable change in modulus. As discussed above, we note that these changes are for a change from 0 to 1/24 defect concentration.

$$\Delta K_{u \to d} = \frac{\Delta a(1-v)}{3V_0(1-2v)} \quad \text{Eq. 6}$$

## 2.5. Relationship between parabolic coefficients and vacancy formation volume $V_{vf,0}$

The parabolic coefficients can also be related to the vacancy formation volume at zero strain (see Appendix C). Here, vacancy formation volume is taken as the change in volume going from the undefected system to the defected system for a single vacancy defect; other texts may call this quantity the vacancy relaxation volume.

This relationship, assuming an isotropic elastic medium, is given in Eq. 7, where $\Delta h = h_d - h_u$ and $V_0$ is the volume of the unstrained, undefected system. The primary term is dependent on $\Delta h$, with an adjustment term dependent on the original undefected $h_u$.

$$V_{vf,0} = 3V_0 \left( \frac{a_d}{a_u} \Delta h + \frac{\Delta a}{a_u} h_u \right) \quad \text{Eq. 7}$$

The vacancy formation volume may also be related to $h_{vf}$ in Eq. 8.

$$V_{vf,0} = 3V_0 \frac{\Delta a}{a_u} h_{vf} \quad \text{Eq. 8}$$

## 2.6. Vacancy relaxation volume

To compare vacancy formation volumes calculated through Eq. 7 with direct calculations, we also calculate a separate vacancy relaxation volume, $V_{vr}$. The $V_{vr}$ have the same conceptual definition as what we call vacancy formation volume, but we call them vacancy relaxation volume because the cells are actually being volume-relaxed. Other texts may differentiate between vacancy formation volume and vacancy relaxation volume on a theoretical level.

For the $V_{vr}$, the starting cubic supercell is defected and then relaxed in volume, relaxed internally, and relaxed in volume again, all at zero strain. The starting cubic volume is then subtracted from the resultant defected cubic volume to produce vacancy relaxation volume. Error bars for $V_{vr}$ values assume a 1Å³ error in each contributing volume. This error estimate is higher than the volume error based solely on calculation convergence, in order to account for the calculation finding and relaxing to a close polymorph (particularly during the internal relaxation step) rather than to a ground state, which can occur in some cases.

## 3. Results

For each system, Fig. 2 shows defected and undefected energy versus strain curves along with a least-squares parabolic fit. Fig. 2 shows the following for these systems:

1. The undefected and defected energy versus strain curves are all generally well approximated by a parabola. Individual supercell energies in these systems deviate by less than 0.05 eV per 40-atom supercell (0.00125 eV/atom) from the parabolic fits.
2. Where the vertex of a defected parabola is notably different from the vertex of the undefected parabola, the defected parabola shows the lowest energy at a slightly positive tensile strain, and a more positive strain than the undefected parabola.
3. The two defected parabolas and the undefected parabola may have the same or different curvatures.
4. Tilt system, B-site cation, and the presence of Sr dopants can all make a difference in the parabolic shapes.

These observations form the foundation for the following discussions.

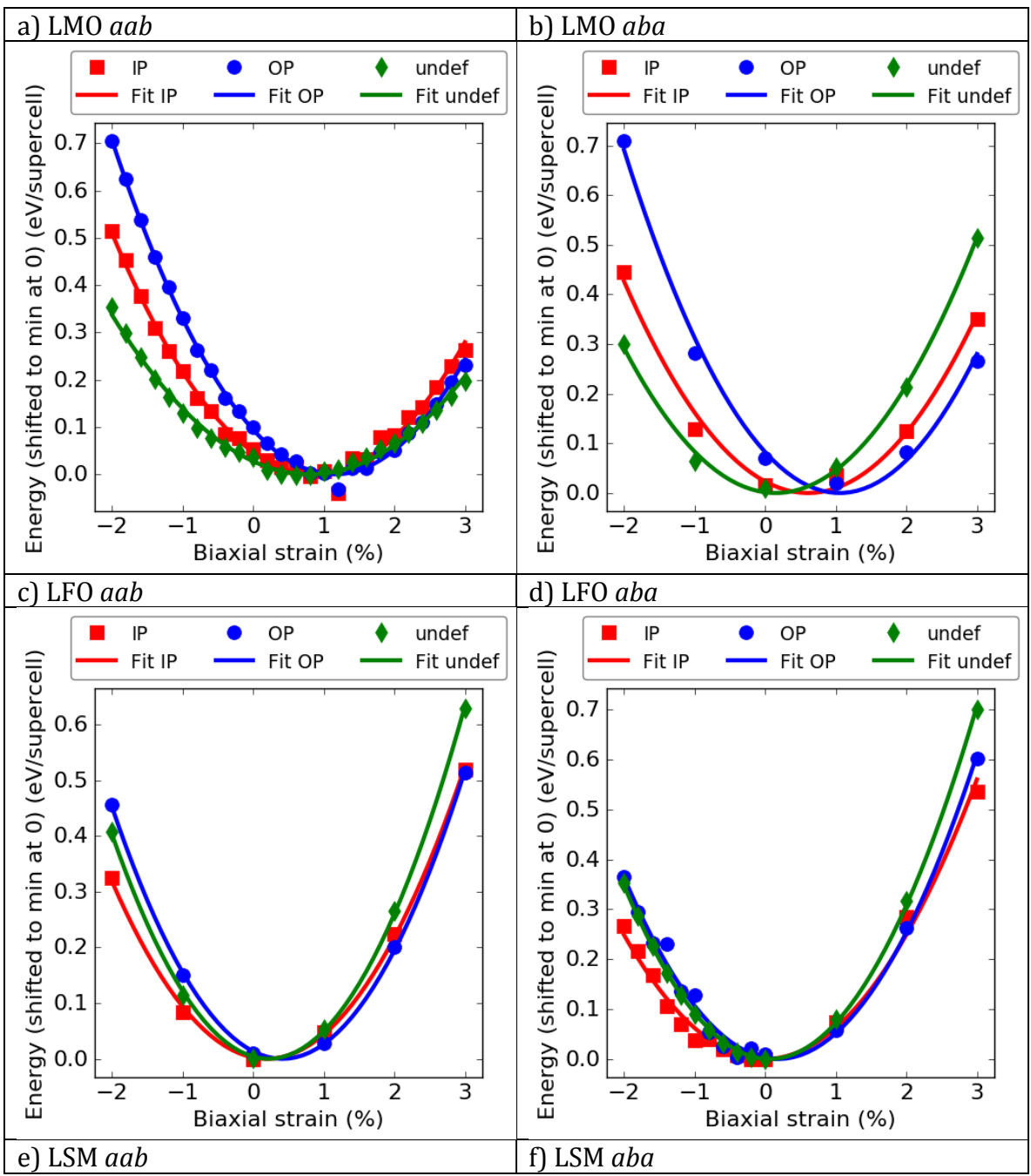

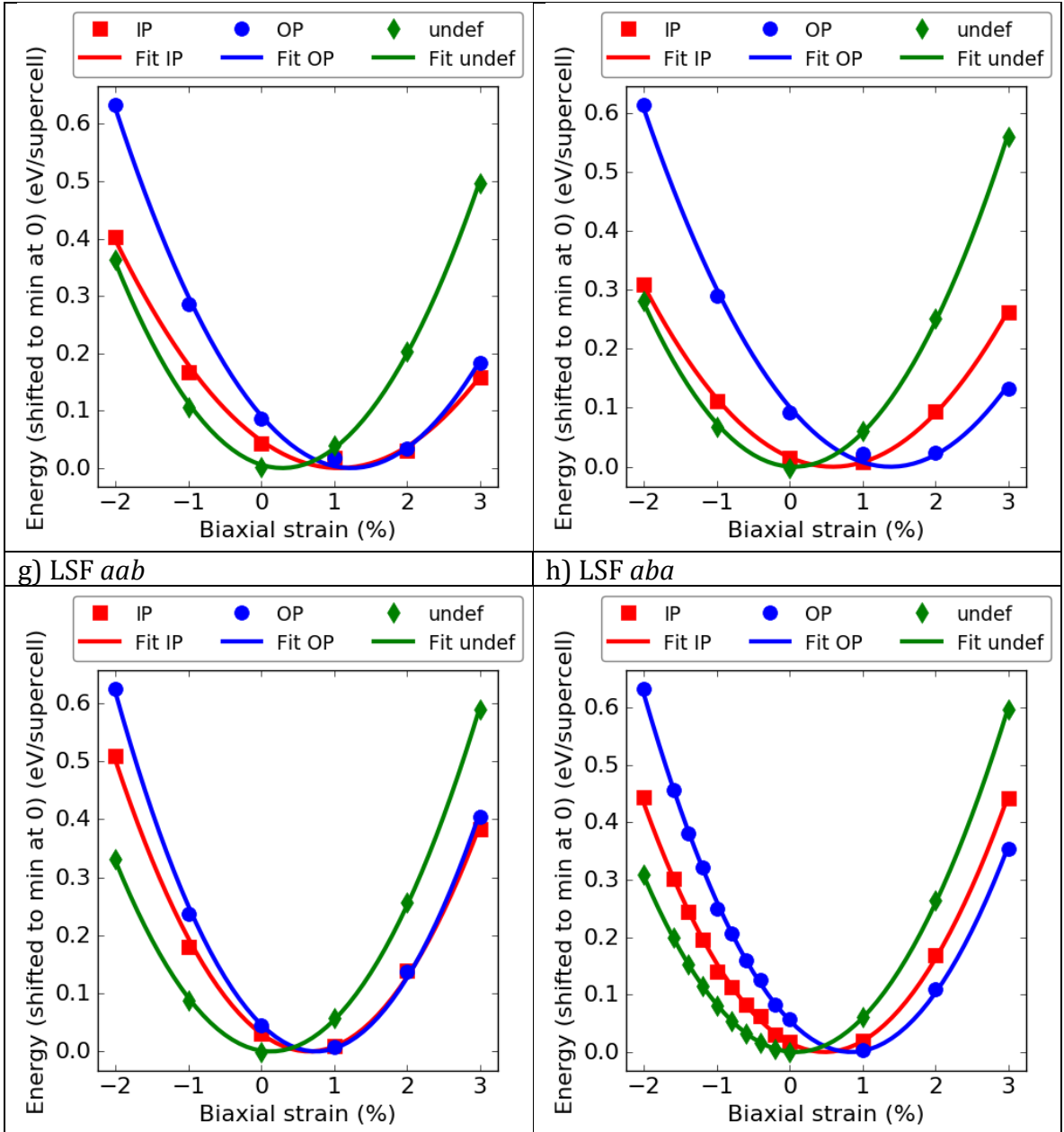

| g) LSF *aab* | h) LSF *aba* |

Fig. 2. Supercell energy versus strain, along with parabolic fits. Curves are shifted to put the minimum value of the fit at zero energy.

Because tilt system can be preserved in thin films,[59] and because Fig. 2 shows that tilt system can have a noticeable effect on the energy versus strain curves, it is potentially important to be sure the system is in the most stable tilt system. Fig. 3 shows the relative stability of the *aba* and *aab* tilt systems, demonstrating that they are nearly equivalent at zero strain, while *aba* is preferred at compressive strains, and *aab* is preferred at tensile strains. This finding is expected because the tilt system makes the $b^+$ axis naturally shorter than the $a^-$ axes[8], so it is expected that the $b^+$ will align along the direction of compression, which is in the xy plane for compression and along the z-axis for tension. The differences in magnitude of the effect between different systems may be due to their different lattice parameters and tilt angles, affecting the amount of distortion. The lower differences for the Sr-doped supercells may have resulted from the body-diagonal distribution of the Sr, which may have stabilized orientation differences. In all cases, the differences were less than 10 meV/atom.

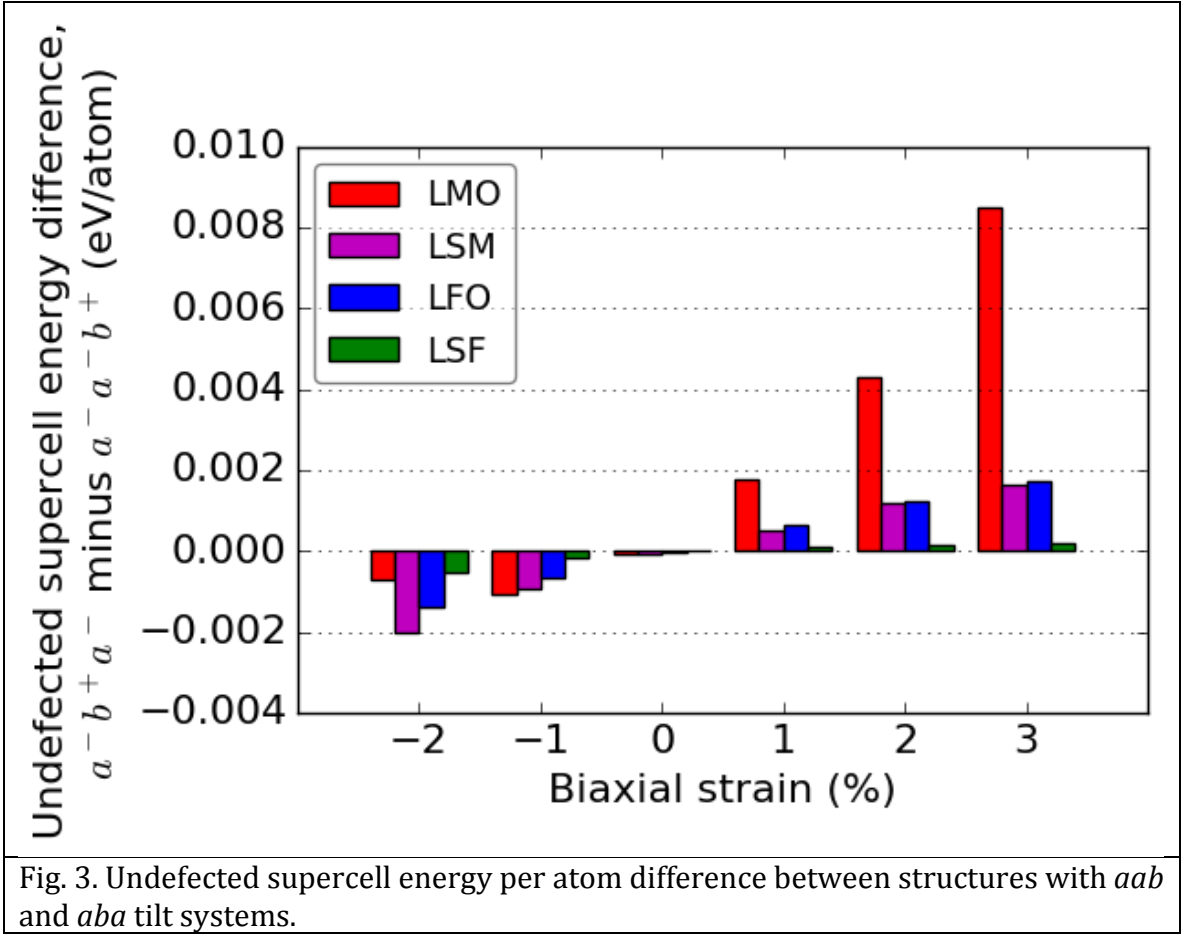

Fig. 3. Undefected supercell energy per atom difference between structures with *aab* and *aba* tilt systems.

Fig. 4 shows the calculated vacancy formation energy versus strain for each system. To the extent that a parabola can represent each supercell energy versus strain, the vacancy formation energy versus strain as calculated by Eq. 1 is then the difference between two parabolas, which is also a parabola. These vacancy formation energy versus strain plots are sensitive to differences in curvature and minimum vertex placement with respect to strain of the defected and undefected supercell energy parabolas. In particular, where the vacancy formation energy with strain is more linear, the defected and undefected energy parabolas have similar curvatures, but displaced minima along the strain axis. Where the vacancy formation energy with strain is more parabolic, the defected and undefected energy parabolas have different curvatures, and the minima are nearer each other along the strain axis.

The effect of strain on vacancy formation energy appears to vary widely between B-site cations and IP and OP vacancies and again between tilt systems. A dashed line connects points for the preferred tilt system at each strain state.

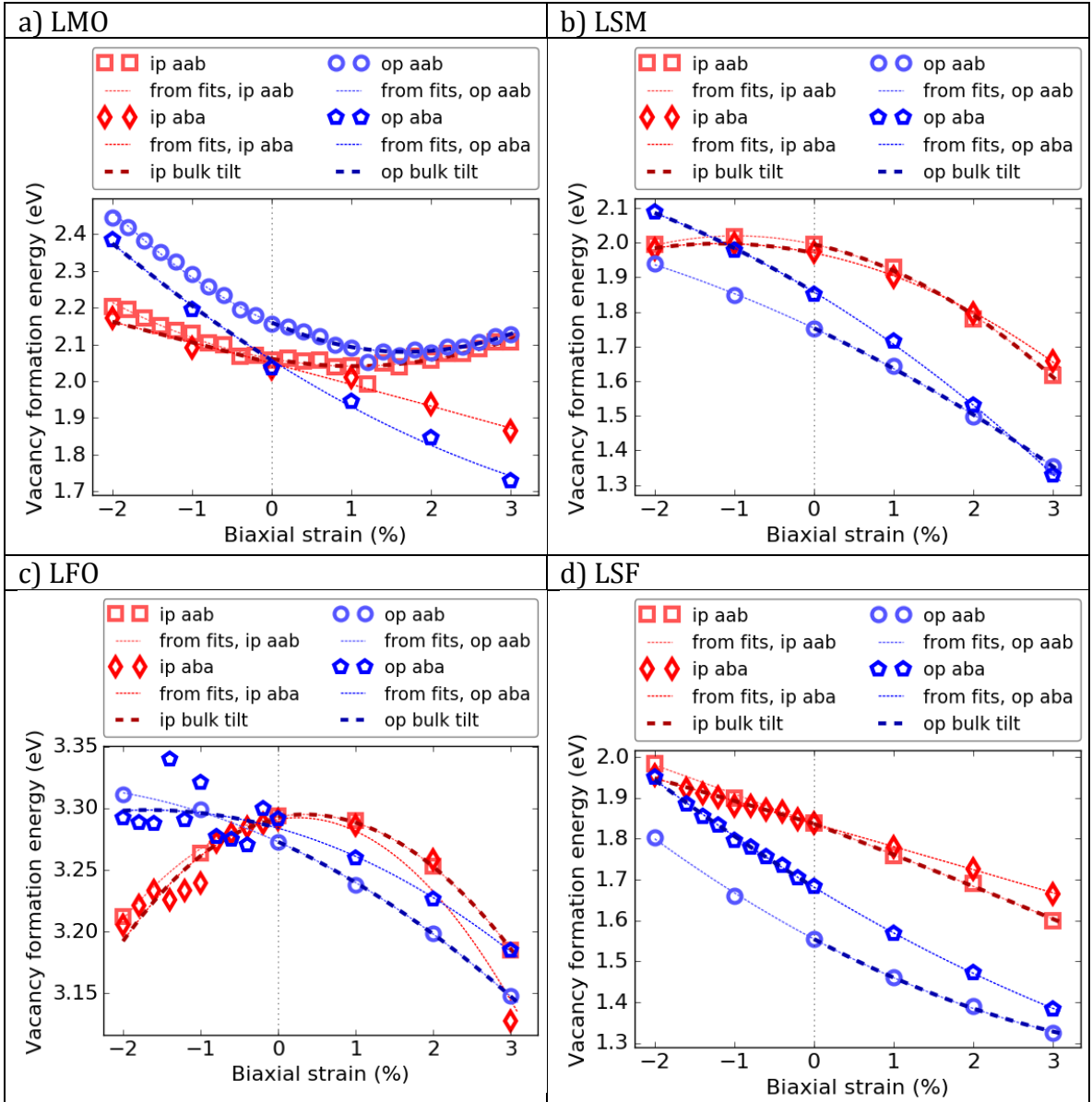

Fig. 4. Vacancy formation energy versus strain, using an oxygen chemical potential for 1173K and 0.1 atm $pO_2$. "Bulk tilt" dashed lines connect points for the lower energy tilt system of the bulk at each strain state. Lines "from fits" indicate the vacancy formation energies resulting from the fitted energy parabolas in Fig. 2.

Fig. 5 plots changes in vacancy formation energy from zero strain for IP and OP vacancies. Fig. 5 shows that tensile strain decreases oxygen vacancy formation energy for most systems (where system includes B-site cation, A-site doping, tilt system, and vacancy type), with the exception of LMO *aab*, while compressive strain has various effects. This asymmetry is due to the fact the vacancies typically have a positive formation volume.

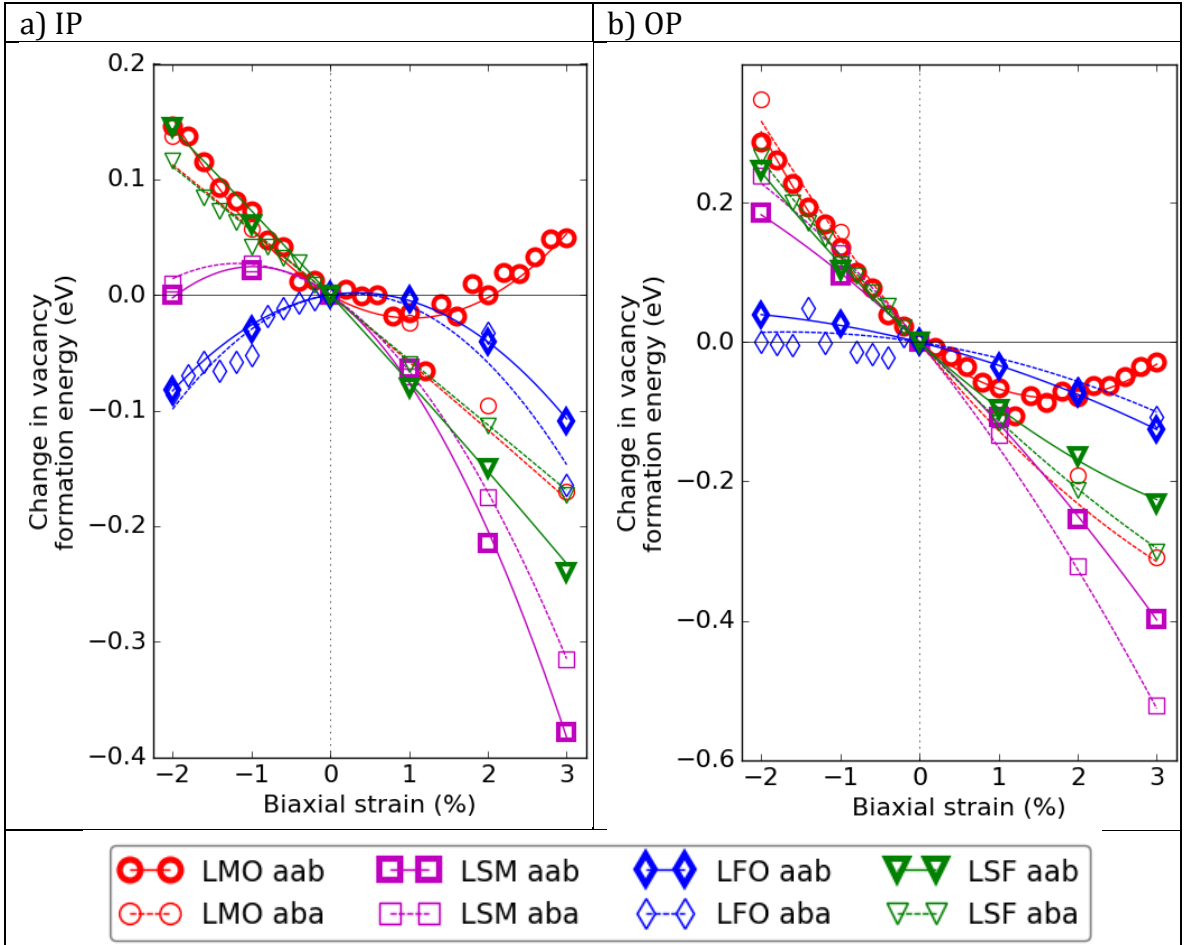

Fig. 5. Changes in vacancy formation energy versus strain for (a) an IP vacancy and (b) an OP vacancy, with fit lines shown.

Fig. 6 plots $\Delta a$ (defined in Eq. 4) for different systems. For many, but not all, systems, Fig. 6 shows negative $\Delta a$. OP $\Delta a$ are all greater (more positive or less negative) than IP $\Delta a$.

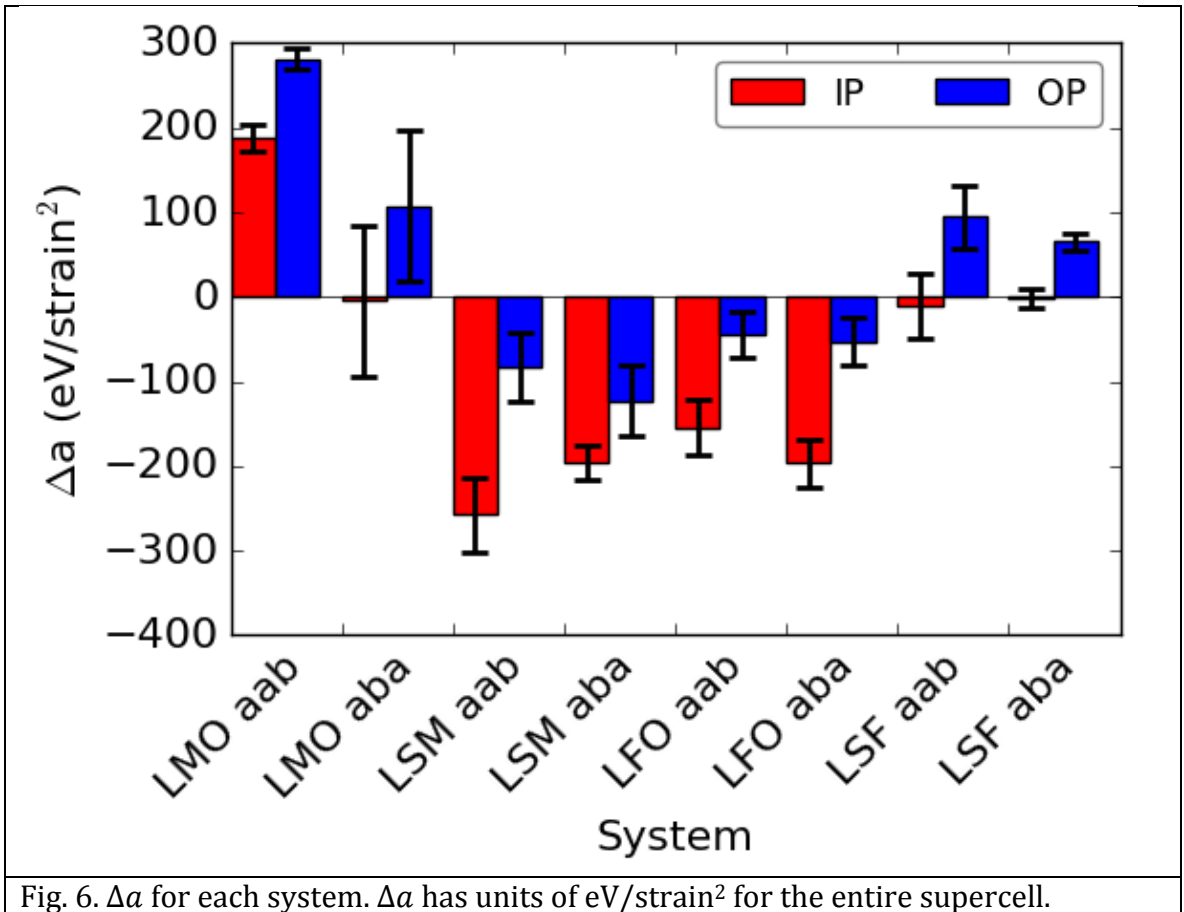

Fig. 6. Δ$a$ for each system. Δ$a$ has units of eV/strain² for the entire supercell.

Fig. 7 shows zero-strain vacancy formation volume $V_{vf,0}$ for different systems, as calculated from Eq. 7. Most values of $V_{vf,0}$ are positive, and those that are not are within 2 Å³ of zero. The OP values are all larger than the IP values.

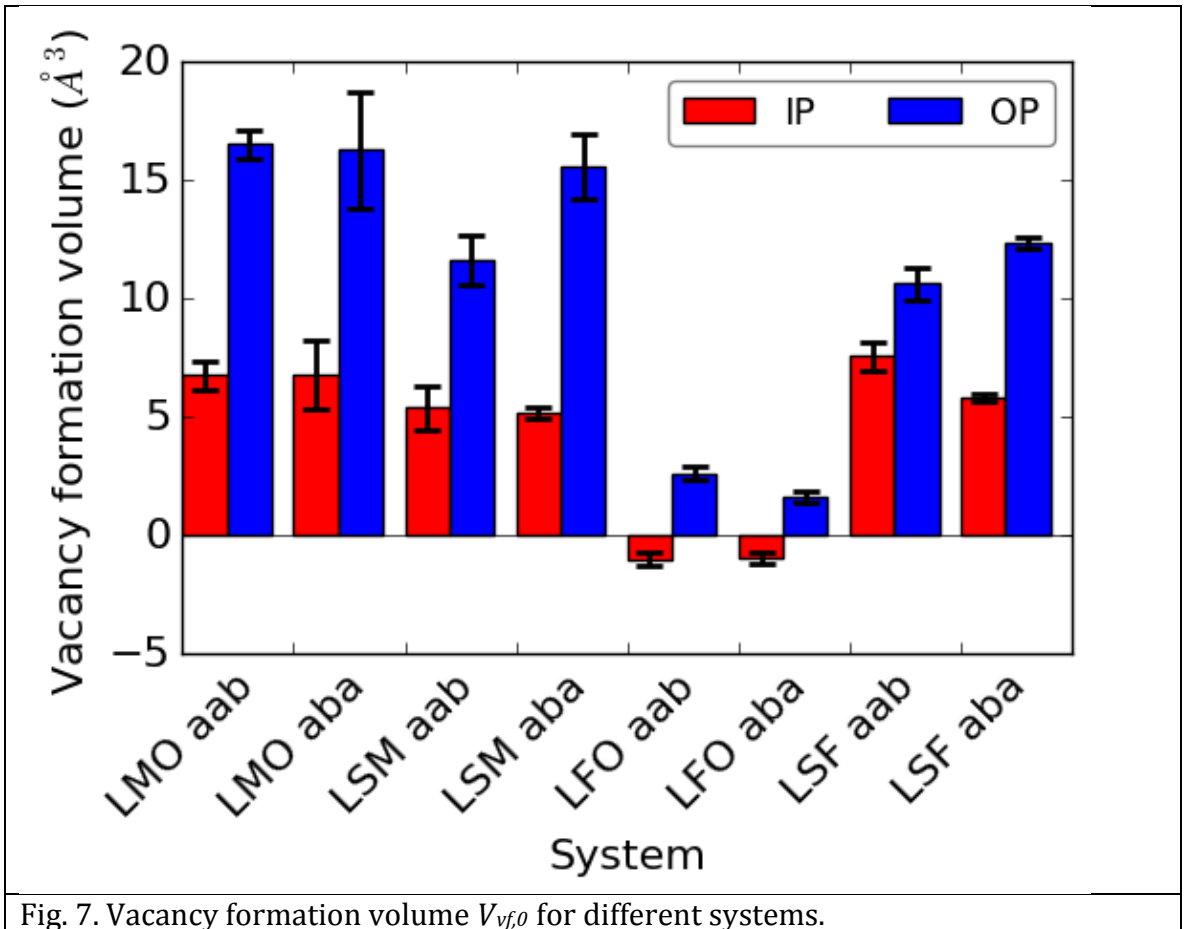

Fig. 7. Vacancy formation volume $V_{vf,0}$ for different systems.

Fig. 8 compares vacancy formation volumes calculated from Eq. 7 with vacancy relaxation volumes $V_{vr}$ calculated through direct relaxation of cubic supercells. The $V_{vr}$ are similar for IP and OP vacancies, and are generally more similar to the IP $V_{vf,0}$ than to the OP $V_{vf,0}$.

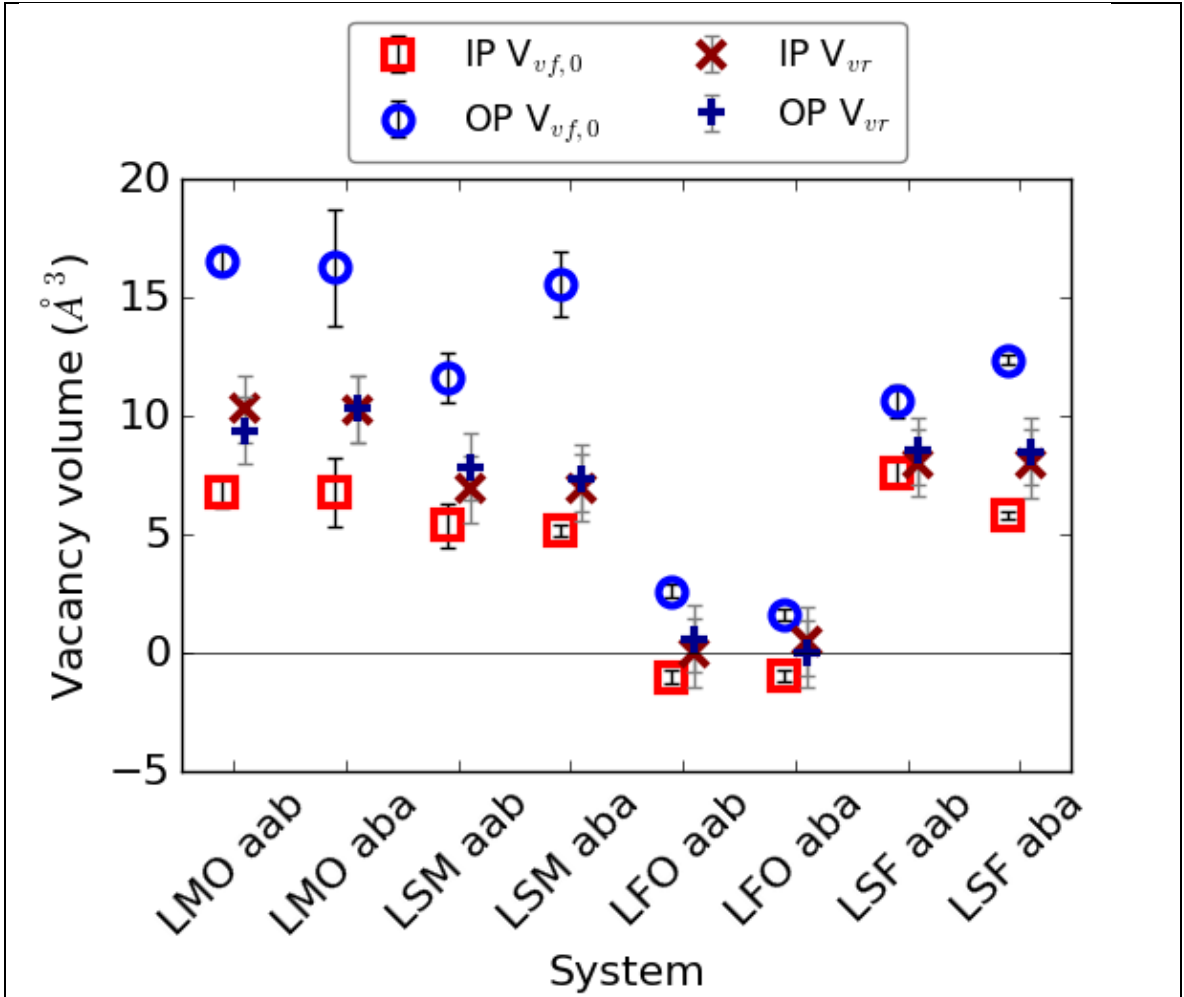

Fig. 8. Vacancy formation volume $V_{vf,0}$ from strained curves in Eq. 7 versus cubic vacancy relaxation volume $V_{vr}$ for different systems. Error bars for vacancy formation volume are given from the errors in the parabolic coefficients. Error bars for vacancy relaxation volume assume a 1Å$^3$ error in each contributing volume (see the Methods section).

## 4. Discussion

### 4.1. Polymorphic curves

Fig. 3 showed that the *aab/aba* systems switch from preferring *aba* to preferring *aab* when transitioning from compressive to tensile strain. This behavior is manifest in Fig. 4 as jumping between vacancy formation energy curves, following the dashed lines. Just as systems may switch low energy bulk tilt systems, they may also switch among different locally stable structures, which we will refer to as polymorphs, within a tilt system. The curves for a property of the system that arise from different structural polymorphs will be denoted as polymorphic curves. An example of a polymorphic curve may be in a single bond, where one curve tracks the progress of a tilt system with smooth variations in each bond angle with strain, while a discontinuity in the progress of a single bond may move the system to a different curve; that second curve may also track the progress of a tilt system with smooth variations in each bond angle with strain, but perhaps with a different range of bond angles for that particular bond.

Taking LMO aab from Fig. 2(a) as an example, even where the supercell energy versus strain curve is fairly smooth and predictable, there are still occasional outliers in the range of 0.05 eV/supercell. The smoothness of most of the curve indicates that the ground states found at each strain, and the ground state progression from strain to strain, are not accidental. However, the outliers also indicate the presence of a close and in some cases slightly more favorable polymorph. Similarly, LFO OP *aba* from Fig. 4(c) is another example, where in the compressive region between -2% and 0% strain we obtain a jagged vacancy formation energy curve which may actually be two or more separate curves. However, Fig. 5(b: OP) shows that the total numeric effect is small compared to the range of changes across all systems.

At intermediate or high temperatures, for example, in the operating range for SOFC materials, local atomic vibrations would likely wash out polymorph differences due to small variations in single bond angles. Therefore, further refining the calculations to identify the polymorphs and obtain complete smooth curves for all of the systems being studied is likely of limited interest. However, in some systems, polymorphic effects may be more severe and further study could be helpful. $(La,Sr)NiO_3$ is given as such a case in the SI Section S1.

### 4.2. Curve shapes

For a vacancy formation energy curve given by Eq. 3, like those shown in Fig. 4, the parabolic coefficients determine the shape of the curve. The change in curvature $\Delta a$ corresponds to an upward concave curve of $E_{vf}(\epsilon)$ if $\Delta a$ is positive, as shown by Fig. 4(a), LMO IP *aab*. In contrast, the downward concave curve of the Fig. 4(c) LFO IP *aab* curve indicates a negative value of $\Delta a$.

A value of $\Delta a$ close to zero indicates a more linear curve, as in Fig. 4(d) LSF IP *aab*, while a larger magnitude value of $\Delta a$ indicates a steeper parabolic curve.

The vertex of the $E_{vf}(\epsilon)$ parabola is given by $(h_{vf}, k_{vf})$ and affects the value of $E_{vf}$ at a particular strain. For example, with a very large positive value of $h_{vf}$ and a negative value of $\Delta a$, $E_{vf}$ will eventually decrease with increasing tensile strain, but $E_{vf}$ may appear to increase at tensile strains below $h_{vf}$, with the decrease only apparent at tensile strains after $h_{vf}$.

The following sections relate these curve shapes to physical measures.

### 4.3. Elastic behavior changes

Fig. 6 shows a negative $\Delta a$ for many systems, indicating a decrease in elastic constants with the defect and a decrease in vacancy formation energy with sufficient elastic strain. The actual strain value at which a decrease starts appearing is related to the value of $h_{vf}$ in Eq. 3. In general OP $\Delta a$ is larger than the IP $\Delta a$, although the origin of this effect is not clear.

### 4.4. Relation between parabolic coefficients and vacancy formation volume

$h$ is the horizontal shift away from zero of the parabolic vertex of a single energy curve. A positive $h$ value indicates that biaxial tensile strain stabilizes the system, and vice versa. Eq. 7 shows that the primary contribution to $V_{vf}$ is related to $\Delta h = h_d - h_u$. Positive $\Delta h$ indicates that additional biaxial strain stabilizes the defect. Therefore, $\Delta h$ is a measure of the vacancy formation volume. $\Delta h$ is expected to be greater than zero, signifying vacancy formation volumes which are typically positive.[19]

Nearly all of the $V_{vf}$ values in Fig. 7 are positive, as expected. The exceptions are within 2 Å$^3$ of zero. The OP vacancy formation volumes are all greater than the IP vacancy formation volumes. The reason for this trend is not clear, but it may have the same source as the OP vs. IP trend in $\Delta a$ noted above.

The $V_{vr}$ in Fig. 8 are also either all positive or close to zero. The closeness of the IP and OP $V_{vr}$ values to each other makes sense because, in the cubic relaxations of the $V_{vr}$, the IP and OP sites are just symmetry-distinct sites, and the small differences are probably due to small local symmetry and internal relaxation differences.

The similarity of the $V_{vr}$ to the IP $V_{vf,0}$ in Fig. 8 lends some credibility to the method of calculating vacancy formation volume in Eq. 7. However, the dissimilarity of the $V_{vr}$ to the OP $V_{vf,0}$ emphasizes that there may be some issue with how the OP defects are treated, perhaps related to our use of a fixed $c$ axis (see Methods section). Further study of the OP vs. IP behavior for both formation volumes and elastic constant changes is warranted but will not be pursued in this work.

### 4.5. Curve shape analogue in migration barriers

While migration barriers were found to decrease almost linearly with strain,[4] a closer examination shows that some of those migration barrier versus strain curves have a small quadratic nature. Applying an analogous model to that developed here for vacancies to migration barriers, the mostly linear curves in the case of migration barrier would indicate either that elastic constants do not change much between the two defected states (the initial state and the transition state), or that migration volume effects are large enough to mask elastic constant effects, or a combination of both. Given that calculated migration volumes are on the order of 5-10 Å³ (Ref. [4, 5]), similar to many of the calculated vacancy formation or relaxation volumes here, the first explanation that the elastic constants do not change very much is more likely. As the migrating oxygen does not break as many bonds or lead to redox of the transition metals, it is reasonable to expect that it would perturb the elastic contstants less than a vacancy.

### 4.6. Dopant association energy

Although not studied here, it is worth noting that the association of vacancies with defects such as dopants or with other vacancies may lower vacancy mobility.[60] Therefore, gains in vacancy migration energy or vacancy formation energy with strain may be offset by stronger defect binding energies with strain.

Rushton et al. found stronger defect binding energies under compression (lower binding energies under tension) in defected-fluorite structured $CeO_2$, and attributed this finding to a lower defect volume per defect cluster component under compression.[61] The magnitude of the decrease (weakened binding) in binding energy found by Rushton et al. was near 150 meV/% strain for a nearest-neighbor oxygen vacancy to a Gd or Y dopant.[61] However, further study is needed to see the extent of these effects in perovskites.

For a combined activation energy for diffusion $E_{act} = E_{mig} + E_{form} + E_{assoc}$, our previous work noted an average of 66 meV/% strain decrease in migration barrier,[4] while this current work shows that while $E_{form}$ generally decreases with tensile strain, its behavior is quite variable. Overall, from the present and literature studies, it is reasonable to expect that tensile strain will decrease $E_{mig}$, likely decrease $E_{form}$, and likely decrease $E_{assoc}$, suggesting that strain can drive significantly enhanced transport.

## 5. Conclusions

Both previous literature and our calculations show mixed results for the direction and magnitude of the change in vacancy formation energy with strain. Here we used a set of consistent calculations to model vacancy formation energy variations with strain effects, and showed that our calculated results may be simply explained in terms of vacancy formation volume and changes in elastic constants between the bulk and defected states. Materials whose elastic constants increase with the introduction of a vacancy may show an increase in vacancy formation energy with large tensile strains, and vice versa. The difference in elastic constants governs the curvature of the vacancy formation energy curve. The actual strain values at which increases and decreases appear are governed by a measure of the vacancy formation volume.

Computations may be more sensitive to small variations in structure, including tilt systems, than would be found in experiments. For example, LMO behaved very differently in these computations depending on how its tilt system was oriented with respect to the strain axes. However, in extended finite temperature experiments, a material may have domains of tilt systems, or may explore multiple tilt systems due to thermal excitations, leading to an averaged behavior. The presence of dopants may also lessen the effect of tilt systems, as shown in Fig. 3.

In general, vacancy formation energies for most systems calculated here decreased with tensile strain. The magnitude of the effect and the onset strain is difficult to generalize because of the significantly different curvatures and horizontal shifts involved in the calculations, but spanned roughly 30-100 meV/% biaxial strain for those systems which showed a decrease in vacancy formation energy. As the goal of this paper was not to rigorously quantify vacancy formation energy with strain for the particular systems, but instead to make sense of the different trend shapes in $H_{vf}$ versus strain, these numbers should not be taken as absolute indicators. Experimental validation of the model is still necessary.

## Supplementary Information
Supplementary information is available and includes a discussion of polymorphic effects on energies, a sample with lower vacancy concentration, an assessment of magnetic ordering effects, a discussion of chemical expansion coefficient, a method for transforming supercells, and a tabulation of strain approaches from the literature.

## Acknowledgments
We would like to acknowledge the NSF Graduate Fellowship Program under Grant No. DGE-0718123 and the UW-Madison Graduate Engineering Research Scholars Program for partial funding of T. Mayeshiba. Computing resources in this work benefitted from the use of the Extreme Science and Engineering Discovery Environment (XSEDE), which is supported by National Science Foundation grant number OCI-1053575, and from the computing resources and assistance of the UW-Madison Center For High Throughput Computing (CHTC) in the Department of Computer Sciences. The CHTC is supported by UW-Madison and the Wisconsin Alumni Research Foundation, and is an active member of the Open Science Grid, which is supported by the National Science Foundation and the U.S. Department of Energy's Office of Science. Support for D. Morgan and the MAST tools applied in this work were provided by the NSF Software Infrastructure for Sustained Innovation (SI2) award No. 1148011.

# Appendix A: Vacancy formation energy versus strain as a parabola

This section derives vacancy formation energy versus strain as a parabola, given parabolic undefected and defected energy versus strain curves.

In Eq. A.1, we approximate vacancy formation enthalpy by vacancy formation energy, an approximation appropriate for zero temperature and atmospheric pressure. We also note that these calculations ignore all thermal excitation effects, except within the oxygen chemical potential.

| $H_{vf} \approx E_{vf}$ | Eq. A.1 |

Eq. A.2 shows the vacancy formation energy calculation for a non charge-compensated oxygen vacancy as the difference between the defected and undefected supercell energies, plus an oxygen chemical potential term, where $\Delta n_O$ for a single vacancy is 1, and $\mu_O$ is the oxygen chemical potential, leading to Eq. 1 in the main text.

| $E_{vf} = E_{\text{VASP, defected}} - E_{\text{VASP, undefected}} + \Delta n_O \mu_O$ | Eq. A.2 |

Eq. A.3 shows Eq. A.2 as a series of vacancy formation energies versus strain, where the oxygen chemical potential is assumed to be unaffected by strain.

| $E_{vf}(\epsilon) = E_{\text{VASP, defected}}(\epsilon) - E_{\text{VASP, undefected}}(\epsilon) + \Delta n_O \mu_O$ | Eq. A.3 |

Eq. A.4 (Eq. 2 in the main text) models each supercell energy curve as a parabola versus strain $\epsilon$, given as a quadratic equation in vertex form, where $a$ gives the curvature, $h$ gives the lateral vertex shift, and $k$ gives the vertical vertex shift.

| $E_{VASP}(\epsilon) = a(\epsilon - h)^2 + k.$ | Eq. A.4 |

Therefore, using subscripts $d$ for defected and $u$ for undefected, Eq. A.5 gives each curve of vacancy formation energy versus strain as the subtraction of two parabolas plus the chemical potential term.

| $E_{vf}(\epsilon) = a_d(\epsilon - h_d)^2 + k_d - a_u(\epsilon - h_u)^2 - k_u + \mu_O$ | Eq. A.5 |

This subtraction results in a third parabola, Eq. A.6.

| $\begin{aligned} H_{vf}(\epsilon) &\approx E_{vf}(\epsilon) \\ &= (a_d - a_u)\epsilon^2 - 2(a_d h_d - a_u h_u)\epsilon \\ &\quad + (a_d h_d^2 - a_u h_u^2) + (k_d - k_u) + \mu_O \end{aligned}$ | Eq. A.6 |

Eq. A.7 gives the change in vacancy formation energy with strain in vertex form, corresponding to Eq. 3 in the main text.

| | |
|---|---|
| $H_{vf}(\epsilon) \approx E_{vf}(\epsilon) = a_{vf}(\epsilon - h_{vf})^2 + k_{vf}$ | Eq. A.7 |

Change constants are defined in Eq. A.8 and Eq. A.9, with equation constants for Eq. A.7 defined in Eq. A.8, Eq. A.10, and Eq. A.11.

| | |
|---|---|
| $a_{vf} = \Delta a = a_d - a_u$ | Eq. A.8 |
| $\Delta h = h_d - h_u$ | Eq. A.9 |
| $h_{vf} = \dfrac{a_d h_d - a_u h_u}{a_d - a_u} = \dfrac{a_d h_d - a_d h_u + a_d h_u - a_u h_u}{\Delta a} = \dfrac{a_d}{\Delta a}\Delta h + h_u$ | Eq. A.10 |
| $k_{vf} = (k_d - k_u) + \mu_O - \dfrac{a_u a_d (h_d - h_u)^2}{(a_d - a_u)}$ | Eq. A.11 |

For reference, Eq. A.12 gives the change in vacancy formation energy as a function of strain.

| | |
|---|---|
| $\Delta H_{vf, 0 \to \epsilon} \approx \Delta E_{vf, 0 \to \epsilon} = (a_d - a_u)\epsilon^2 - 2(a_d h_d - a_u h_u)\epsilon$ $= \Delta a \epsilon^2 - 2(a_d h_d - a_u h_u)\epsilon$ | Eq. A.12 |

# Appendix B: Elastic constants and $\Delta a$

This section derives the relationship between elastic constants and $\Delta a$. These derivations follow from Chapter 4 of Boresi, et al.,[62] and specific equations used are referenced below.

First, we relate strain energy density *u* to elastic constants and strain, assuming plane stress (biaxial strain with out-of-plane relaxation resulting in stress along x and y, and no stress along z) in a homogeneous medium with no shear strain and no shear stress.

Eq. B.1 (Boresi Eq. 4-4.7) gives a general formula for strain energy density, with subscripts in Eq. B.2.

| | |
|---|---|
| $u = \dfrac{1}{2} C_{\alpha\beta} \epsilon_\alpha \epsilon_\beta$ | Eq. B.1 |
| $\alpha, \beta \in \{1,2,3,4,5,6\}$ | Eq. B.2 |

Eq. B.3 shows a decreased subscript list for the case without shear strain (all calculation supercells are cubic or tetragonal), giving the expanded expression for strain energy density in Eq. B.4.

| | |
|---|---|
| $\epsilon_4 = \epsilon_5 = \epsilon_6 = 0 \rightarrow \alpha, \beta \in \{1,2,3\}$ | Eq. B.3 |
| $u_{no\ shear\ strain}$ $= \frac{1}{2}C_{11}\epsilon_1\epsilon_1 + \frac{1}{2}C_{12}\epsilon_1\epsilon_2 + \frac{1}{2}C_{13}\epsilon_1\epsilon_3 + \frac{1}{2}C_{21}\epsilon_2\epsilon_1$ $+ \frac{1}{2}C_{22}\epsilon_2\epsilon_2 + \frac{1}{2}C_{23}\epsilon_2\epsilon_3 + \frac{1}{2}C_{31}\epsilon_3\epsilon_1 + \frac{1}{2}C_{32}\epsilon_3\epsilon_2$ $+ \frac{1}{2}C_{33}\epsilon_3\epsilon_3$ | Eq. B.4 |

For a homogenous medium, the stiffness tensor is symmetric (Eq. B.5), giving a simplified equation for strain energy density in Eq. B.6.

| | |
|---|---|
| $C_{12} = C_{21}, C_{13} = C_{31}, C_{23} = C_{32}$ | Eq. B.5 |
| $u = \frac{1}{2}C_{11}\epsilon_1\epsilon_1 + C_{12}\epsilon_1\epsilon_2 + C_{13}\epsilon_1\epsilon_3 + \frac{1}{2}C_{22}\epsilon_2\epsilon_2 + C_{23}\epsilon_2\epsilon_3 + \frac{1}{2}C_{33}\epsilon_3\epsilon_3$ | Eq. B.6 |

Applying the biaxial strain condition in Eq. B.7 and Eq. B.8 gives Eq. B.9.

| | |
|---|---|
| $\epsilon_1 = \epsilon_2 = \epsilon_x = \epsilon_y$ | Eq. B.7 |
| $\epsilon_3 = \epsilon_z$ | Eq. B.8 |
| $u = \left(\frac{1}{2}C_{11} + C_{12} + \frac{1}{2}C_{22}\right)\epsilon_x^2 + (C_{13} + C_{23})\epsilon_x\epsilon_z + \frac{1}{2}C_{33}\epsilon_z^2$ | Eq. B.9 |

To obtain an expression solely in $\epsilon_x$, Eq. B.10 (Boresi Eq. 4-4.2) relates stress to strain and elastic constants, and the conditions of plane stress, no shear stress, and no shear strain give Eq. B.11. Thus, expansion of Eq. B.10 for $\sigma_3$ (and also taking into account a symmetric stiffness tensor as above) gives Eq. B.12, which can be rearranged into Eq. B.13.

| | |
|---|---|
| $\sigma_\alpha = C_{\alpha\beta}\epsilon_\beta$ | Eq. B.10 |
| $\sigma_3 = 0, \sigma_4 = \sigma_5 = \sigma_6 = \epsilon_4 = \epsilon_5 = \epsilon_6 = 0$ | Eq. B.11 |
| $\sigma_3 = 0 = C_{31}\epsilon_1 + C_{32}\epsilon_2 + C_{33}\epsilon_3 = C_{13}\epsilon_x + C_{23}\epsilon_x + C_{33}\epsilon_z$ | Eq. B.12 |
| $\epsilon_z = \frac{-(C_{13} + C_{23})\epsilon_x}{C_{33}}$ | Eq. B.13 |

Substituting Eq. B.13 into Eq. B.9 gives Eq. B.14, with strain energy density in terms of $\epsilon_x$. Algebraic simplification of Eq. B.14 gives Eq. B.15.

| | |
|---|---|
| $u = \left(\frac{1}{2}C_{11} + C_{12} + \frac{1}{2}C_{22}\right)\epsilon_x^2 + (C_{13} + C_{23})\epsilon_x \left(\frac{-(C_{13} + C_{23})\epsilon_x}{C_{33}}\right)$ $+ \frac{1}{2}C_{33}\left(\frac{-(C_{13} + C_{23})\epsilon_x}{C_{33}}\right)^2$ | Eq. B.14 |
| $u = \left(\frac{1}{2}C_{11} + C_{12} + \frac{1}{2}C_{22} - \frac{1}{2}\left(\frac{(C_{13} + C_{23})^2}{C_{33}}\right)\right)\epsilon_x^2$ | Eq. B.15 |

Using additional approximations for an isotropic medium gives Eq. B.16, Eq. B.17, and Eq. B.18 (Boresi Eq. 4-6.2 and surrounding text), defining the Lamé constants $\lambda$ and $G$. Eq. B.18 transforms into Eq. B.19, which can then be substituted into Eq. B.17 to give Eq. B.20.

| | |
|---|---|
| $C_{23} = C_{13} = C_{12} = C_2 = \lambda$ | Eq. B.16 |
| $C_{11} = C_{22} = C_{33} = C_1$ | Eq. B.17 |
| $G = \frac{C_1 - C_2}{2} = \frac{C_1 - \lambda}{2}$ | Eq. B.18 |
| $C_1 = 2G + \lambda$ | Eq. B.19 |
| $C_{11} = C_{22} = C_{33} = 2G + \lambda$ | Eq. B.20 |

Substituting Eq. B.16 and Eq. B.20 into Eq. B.15 gives Eq. B.21, which simplifies into Eq. B.22.

| | |
|---|---|
| $u = \left(\frac{1}{2}(2G + \lambda) + \lambda + \frac{1}{2}(2G + \lambda) - \frac{1}{2}\left(\frac{(2\lambda)^2}{(2G + \lambda)}\right)\right)\epsilon_x^2$ | Eq. B.21 |
| $u = \left(\frac{2G(2G + 3\lambda)}{2G + \lambda}\right)\epsilon_x^2$ | Eq. B.22 |

Now, we will relate the parabolic curvature coefficient *a* to elastic constants. We start with the parabolic energy curve as defined in Eq. B.23, where $U$ is strain energy, or total supercell energy shifted so that it is zero at zero strain. Since an elastic comparison looks only at changes in the elastic response shown by the curvature, we then ignore vertex position, setting *h=0* and *k=0*, giving Eq. B.24.

| | |
|---|---|
| $U = a(\epsilon_x - h)^2 + k$ | Eq. B.23 |
| $U = a\epsilon_x^2$ | Eq. B.24 |

Eq. B.25 gives the relationship between strain energy and strain energy density, with volume at strain defined in Eq. B.26.

| | |
|---|---|
| $U(\epsilon_x) = u_{\epsilon_x} V_{\epsilon_x}$ | Eq. B.25 |
| $V_{\epsilon_x} = V_0 + f(\epsilon_x)$ | Eq. B.26 |

Substituting Eq. B.24 into Eq. B.25 gives Eq. B.27.

| $a\epsilon_x^2 = u_{\epsilon_x} V_{\epsilon_x}$ | Eq. B.27 |
|---|---|

Substituting Eq. B.22 into Eq. B.27 gives Eq. B.28.

| $a\epsilon_x^2 = V_{\epsilon_x} \left( \dfrac{2G(2G + 3\lambda)}{2G + \lambda} \right) \epsilon_x^2$ | Eq. B.28 |
|---|---|

Canceling out the strain terms from both sides and considering all elastic constants, and therefore also volume, at zero strain gives Eq. B.29.

| $a = V_0 \left( \dfrac{2G(2G + 3\lambda)}{2G + \lambda} \right)$ | Eq. B.29 |
|---|---|

Elastic constant transformations can then relate $a$ to bulk modulus $K$ and Poisson's ratio $\nu$.

Eq. B.30, Eq. B.31, and Eq. B.32 (all adapted from Boresi Eq. 4-6.9a) give transformation equations for elastic constants.

| $\lambda = \dfrac{\nu Y}{(1 + \nu)(1 - 2\nu)}$ | Eq. B.30 |
|---|---|
| $G = \dfrac{Y}{2(1 + \nu)}$ | Eq. B.31 |
| $Y = 3K(1 - 2\nu)$ | Eq. B.32 |

Substituting Eq. B.32 into Eq. B.30 and into Eq. B.31 gives Eq. B.33 and Eq. B.34, respectively.

| $\lambda = \dfrac{3K\nu}{(1 + \nu)}$ | Eq. B.33 |
|---|---|
| $G = \dfrac{3K(1 - 2\nu)}{2(1 + \nu)}; 2G = \dfrac{3K(1 - 2\nu)}{(1 + \nu)}$ | Eq. B.34 |

Combining Eq. B.33 and Eq. B.34 gives Eq. B.35 and Eq. B.36, and combining Eq. B.34 and Eq. B.36 gives Eq. B.37.

| $2G + 3\lambda = \dfrac{3K(1 - 2\nu)}{(1 + \nu)} + \dfrac{9K\nu}{(1 + \nu)} = \dfrac{3K - 6K\nu + 9K\nu}{(1 + \nu)} = \dfrac{3K(1 + \nu)}{(1 + \nu)} = 3K$ | Eq. B.35 |
|---|---|
| $2G + \lambda = \dfrac{3K(1 - 2\nu)}{(1 + \nu)} + \dfrac{3K\nu}{(1 + \nu)} = \dfrac{3K - 6K\nu + 3K\nu}{(1 + \nu)} = \dfrac{3K(1 - \nu)}{(1 + \nu)}$ | Eq. B.36 |
| $\dfrac{2G}{2G + \lambda} = \dfrac{3K(1 - 2\nu)}{(1 + \nu)} * \dfrac{(1 + \nu)}{3K(1 - \nu)} = \dfrac{(1 - 2\nu)}{(1 - \nu)}$ | Eq. B.37 |

Finally, substituting Eq. B.35 and Eq. B.37 into Eq. B.29 gives Eq. B.38, which can then be transformed into Eq. B.39 to relate bulk modulus $K$ to parabolic curvature coefficient $a$. Eq. B.40 expresses Eq. B.39 as a change in values from the undefected state to the defected state.

| | |
|---|---|
| $a = \dfrac{3V_0 K(1-2\nu)}{(1-\nu)}$ | Eq. B.38 |
| $K = \dfrac{a(1-\nu)}{3V_0(1-2\nu)}$ | Eq. B.39 |
| $\Delta K_{u \to d} = \dfrac{\Delta a(1-\nu)}{3V_0(1-2\nu)}$ | Eq. B.40 |

## Appendix C: Parabolic coefficients and $V_{vf}$

This section derives the relationship between the parabolic coefficients of $E_{vf}(\epsilon)$, $E_d(\epsilon)$, and $E_u(\epsilon)$ and the zero strain vacancy formation volume $V_{vf,0}$ using an isotropic elastic approximation.

From Schichtel et al.,[63] Eq. C.1 through Eq. C.3 define isotropic pressure near an interface with biaxial strain, assuming an elastic medium.

| | |
|---|---|
| $P = \dfrac{1}{3}(\sigma_{xx} + \sigma_{yy} + \sigma_{zz})$ | Eq. C.1 |
| $\sigma_{xx} = \sigma_{yy} = \dfrac{-Y}{1-\nu}\epsilon_{12}; \sigma_{zz} = 0$ | Eq. C.2 |
| $P = -\dfrac{2}{3}\dfrac{Y}{(1-\nu)}\epsilon_{12} = -\dfrac{2}{3}\dfrac{Y}{(1-\nu)}\epsilon$ | Eq. C.3 |

Eq. C.4 shows vacancy formation enthalpy.

| | |
|---|---|
| $H_{vf} = U_{vf} + PV_{vf} + \Delta n_O \mu_O$ | Eq. C.4 |

Eq. C.5 defines vacancy formation volume.

| | |
|---|---|
| $V_{vf} = \left(\dfrac{\partial H_{vf}}{\partial P}\right)_{S,\mu_O} = \left(\dfrac{\partial H_{vf}}{\partial \epsilon}\right)_{S,\mu_O}\left(\dfrac{\partial \epsilon}{\partial P}\right)_{S,\mu_O} = \left(\dfrac{\partial H_{vf}}{\partial \epsilon}\right) \Big/ \left(\dfrac{\partial P}{\partial \epsilon}\right)$ | Eq. C.5 |

Eq. C.6 applies the zero strain condition to the vacancy formation volume definition of Eq. C.5.

| $$V_{vf,0} = \left(\left.\frac{\partial H_{vf}}{\partial \epsilon}\right|_{\epsilon=0}\right) \Big/ \left(\left.\frac{\partial P}{\partial \epsilon}\right|_{\epsilon=0}\right)$$ | Eq. C.6 |
|---|---|

Substituting in the strain derivatives of Eq. A.6 and Eq. C.3 into Eq. C.6 gives Eq. C.7, which is then simplified in Eq. C.8 and Eq. C.9.

| $$V_{vf,0} = \frac{[2(a_d - a_u)\epsilon - 2(a_d h_d - a_u h_u)]_{\epsilon=0}}{\left[-\frac{2}{3}\frac{Y}{(1-\nu)}\right]_{\epsilon=0}}$$ | Eq. C.7 |
|---|---|
| $$V_{vf,0} = \frac{-3(1-\nu)[(a_d - a_u)*0 - (a_d h_d - a_u h_u)]}{Y}$$ | Eq. C.8 |
| $$V_{vf,0} = \frac{3(1-\nu)(a_d h_d - a_u h_u)}{Y}$$ | Eq. C.9 |

Eq. C.10 substitutes Young's modulus with an equivalent isotropic expression in bulk modulus and Poisson's ratio.

| $$V_{vf,0} = \frac{3(1-\nu)(a_d h_d - a_u h_u)}{3K(1-2\nu)}$$ | Eq. C.10 |
|---|---|

Substituting Eq. B.3.15 into Eq. C.10, using the undefected value $a_u$ to correspond with the undefected bulk modulus, gives Eq. C.11.

| $$V_{vf,0} = \frac{3V_0(a_d h_d - a_u h_u)}{a_u}$$ | Eq. C.11 |
|---|---|

Finally, the subsequent equations transform Eq. C.11 into Eq. C.14, which shows the relationship between $V_{vf,0}$ and $\Delta h$, having both a primary term dependent on $\Delta h$ and an adjustment term dependent on the original undefected $h_u$.

| $$V_{vf,0} = \frac{3V_0[a_d h_d - a_d h_u + a_d h_u - a_u h_u]}{a_u}$$ | Eq. C.12 |
|---|---|
| $$V_{vf,0} = \frac{3V_0[a_d(h_d - h_u) + h_u(a_d - a_u)]}{a_u}$$ | Eq. C.13 |
| $$V_{vf,0} = 3V_0\left(\frac{a_d}{a_u}\Delta h + \frac{\Delta a}{a_u}h_u\right)$$ | Eq. C.14 |

Eq. C.14 may also be related to $h_{vf}$ by using Eq. A.10, giving Eq. C.15.

| $$V_{vf,0} = 3V_0\frac{\Delta a}{a_u}h_{vf}$$ | Eq. C.15 |
|---|---|

SUPPLEMENTARY INFORMATION
for
**Strain effects on oxygen vacancy formation energy in perovskites**

Tam Mayeshiba, Dane Morgan

# S1. Polymorphs and energy vs. strain discontinuities in LSN

This appendix discusses discontinuities and polymorphs in $La_{0.75}Sr_{0.25}NiO_3$ (LSN), which complicated the resulting energy curves to such an extent we excluded them from the main paper examples. Fig. S1.1 shows energy versus strain curves for IP and OP LSN calculated by the same approach described in the Methods section, which curves show polymorphs that vary in structure and in magnetic moment.

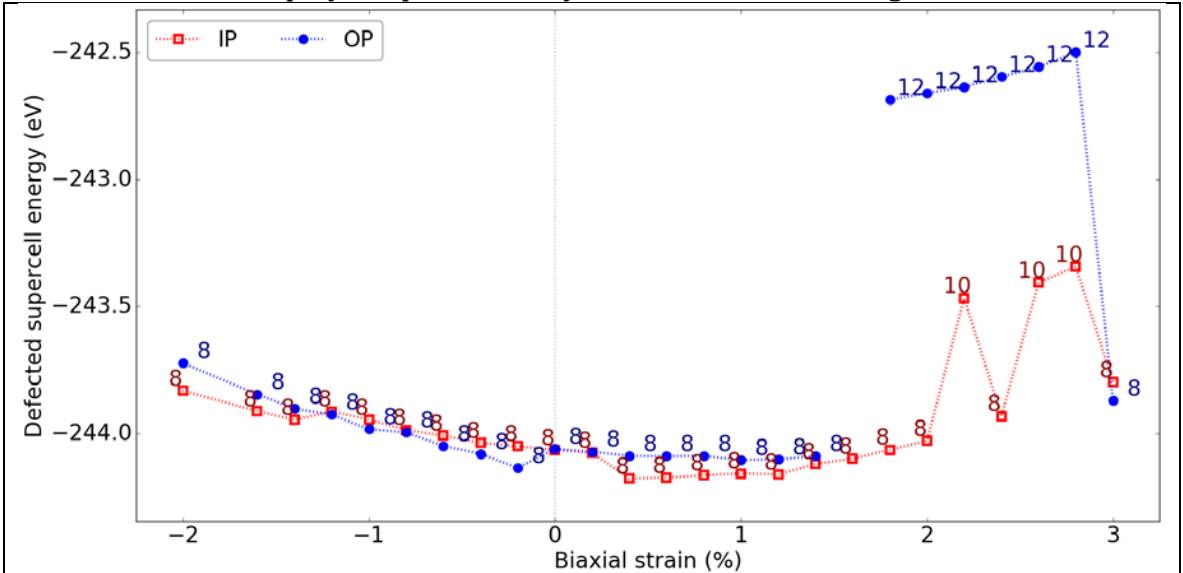

Fig. S1.1. Energy versus strain curves for IP (red open squares) and OP (blue closed circles). Supercell magnetic moments obtained for each point are also labeled.

Changes in energy associated with a magnetic difference are easy to identify from the moments, and are in general quite large (>0.5 eV/supercell). Smaller discontinuities may indicate structural polymorphs, originating for example in a discontinuous change in tilt system. An example is LSN OP, where the raw energies show a discontinuity between -0.2% and 0% strain with an energy difference of almost 100 meV, and the tilt system also shows a discontinuity.

Fig. S1.2 shows images for the LSN OP discontinuity between -0.2% and 0% strain. (Images were rendered using VESTA.[1]) The calculations for all four images in Fig. S1.2 have supercell magnetic moments of 8 $\mu_B$. Images are all from the same view, looking from the origin along the *b*-axis in a right-hand coordinate system. Looking at the oxygen position between the top and bottom right-hand octahedra in the images, there are two non-overlapping oxygen atoms visible at -0.4% strain, 0 strain, and 0.2% strain. However, these oxygen atoms overlap at -0.2% strain, creating a discontinuity when progressing from -0.4% to -0.2% to 0 strain. The overlap or non-overlap is due to different B-O-B bond angles, and is an example of a polymorphic discontinuity.

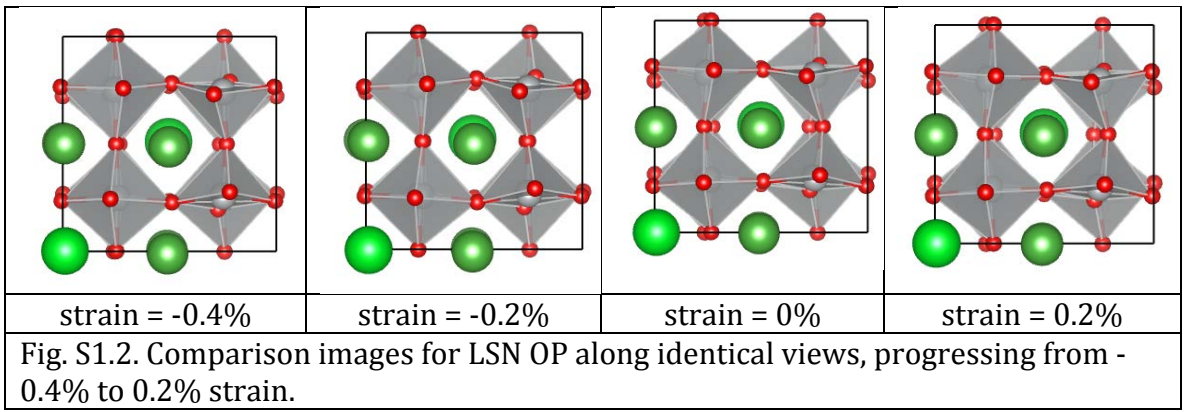

| strain = -0.4% | strain = -0.2% | strain = 0% | strain = 0.2% |

Fig. S1.2. Comparison images for LSN OP along identical views, progressing from -0.4% to 0.2% strain.

A more obvious discontinuous magnetic moment evaluation may be accompanied by a less-obvious discontinuous structure change. The 6 anomalous points in LSN OP have supercell magnetic moments of 12 $\mu_B$ and a supercell energy jump of around 1.4 eV, while the 3 anomalous points in LSN IP have supercell magnetic moments of 10 $\mu_B$ and a supercell energy jump of around 0.6 eV. Fig. S1.3 shows example cells of LSN IP, looking down the *c*-axis to the origin. The upper left A-site cation position (in green) shows two overlapping cations at 2% strain, then two non-overlapping cations at 2.2% strain, then two overlapping cations again at 2.4% strain. Similarly, the oxygen position between the right-hand top and bottom octahedra in the image shows two non-overlapping oxygen ions, then two overlapping ions, then two non-overlapping oxygen ions. Both of these differences account for a non-smooth progression in structure between 2% and 2.4% strain. Forcing the magnetic moment to stay fixed during relaxation may remove this particular structural discontinuity, but we show below that this change is not enough to yield stable trends with strain.

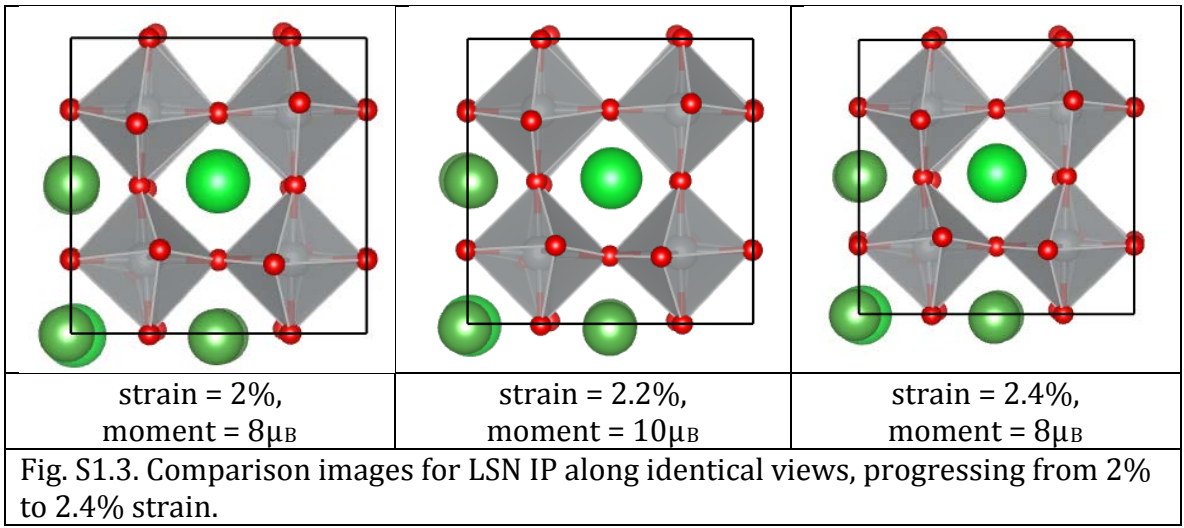

| strain = 2%, moment = 8μ<sub>B</sub> | strain = 2.2%, moment = 10μ<sub>B</sub> | strain = 2.4%, moment = 8μ<sub>B</sub> |

Fig. S1.3. Comparison images for LSN IP along identical views, progressing from 2% to 2.4% strain.

Fig. S1.4 shows an attempt to remove structural discontinuities by restricting the supercell magnetic moment to the value that gives the lowest undefected cell energy (6μ$_B$) and by stepping from -1.4% strain. That is, the supercell coordinates from -1.4% strain were used for -1.2% strain, the resulting coordinates from -1.2% strain were used for -1.0% strain, and so on. The results are compared to our standard approach as described in the Methods section. The revised approach does remove the discontinuity between -1.4% and -1.2% strain, but at approximately 0.2% strain, the energy curve appears to climb up a different polymorphic parabola, higher in energy than the parabola obtained by our standard approach.

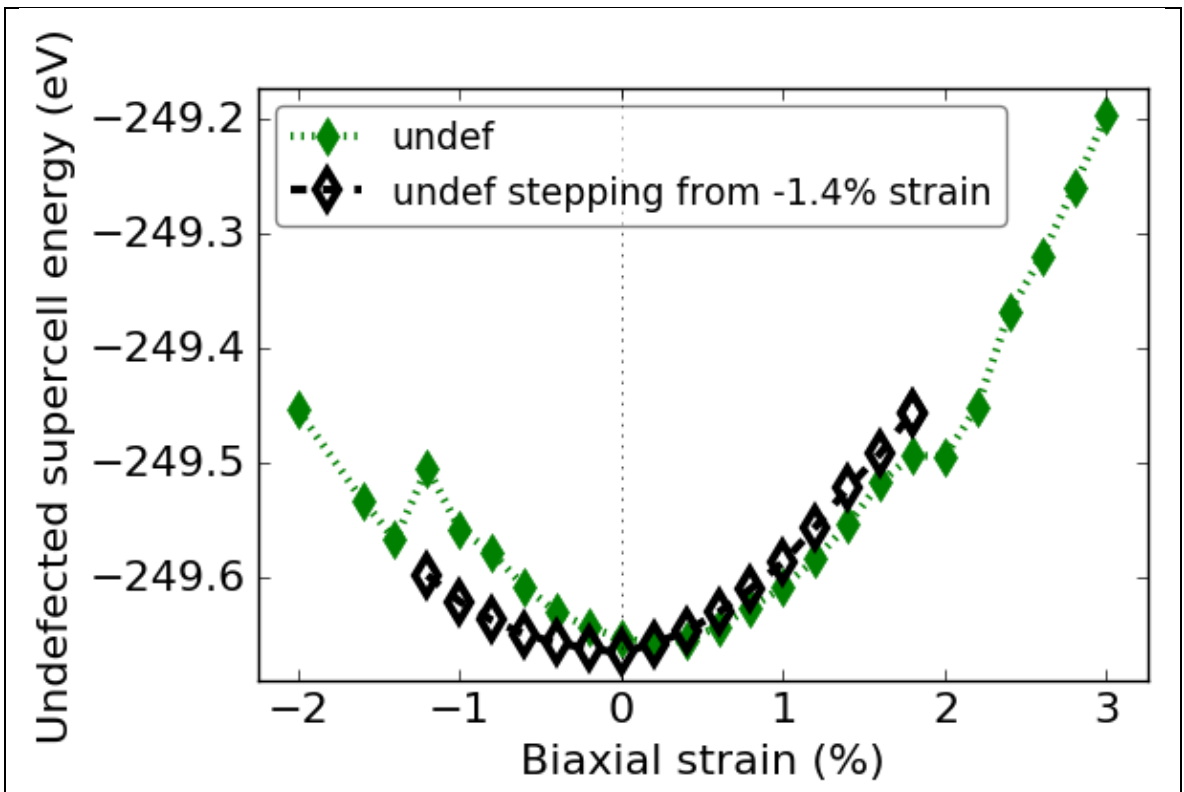

Fig. S1.4. Undefected supercell energy versus strain, with and without the stepping procedure.

Fig. S1.5 shows that fixing the magnetic moment can remove magnetic moment discontinuities. Fig. S1.5 also shows the results of stepping corrections, starting from the stepped undefected supercell at each strain state, in the correction described by Fig. S1.5. While stepping can remove some discontinuities, it is still not effective at removing all discontinuities, especially as the stepping continues further from the compressive regime where it started, into the tensile regime. Stepping and fixing magnetic moments can remove large discontinuities, but discontinuities on the order of 0.1 to 0.2 eV in supercell energy remain.

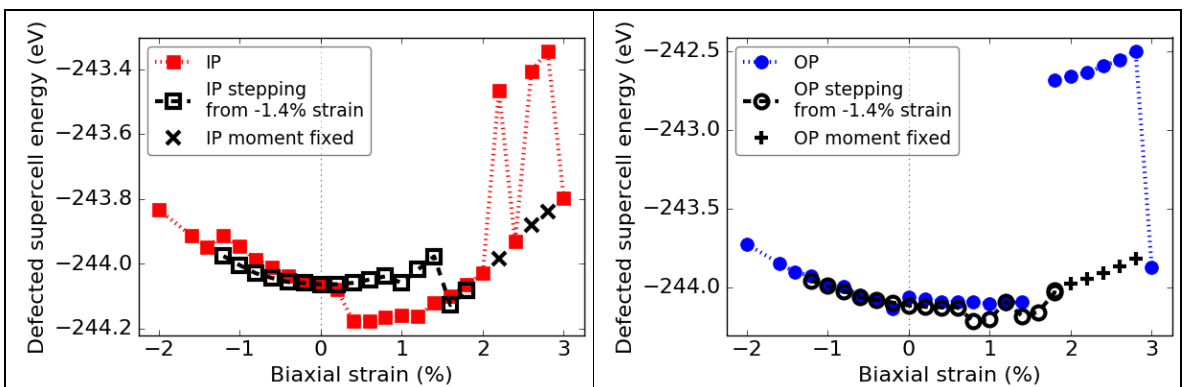

Fig. S1.5. Defected supercell energies with strain, with and without the stepping procedure and fixing magnetic moments.

These results demonstrate that finding the stable structure at each strain may be challenging for some systems, e.g. LSN. We believe that the conclusion in the main text of parabolic vacancy formation energy versus strain is still valid for such systems, but would apply to each polymorph independently. However, demonstrating this relationship can be difficult due to the inability to constrain the system easily to one polymorph, as we found in LSN. Furthermore, experimental results for systems with many polymorphs, especially polymorphs that become more or less preferred at different strain states, may show overall behavior that is a disjointed combination of many parabolas. This is similar to the behavior that is expected due to changing tilt system (see the Results section).

## S2. Lower vacancy concentration

This section shows results for calculations on a single system (LMO) at a lower vacancy concentration, with a single oxygen vacancy in a 4x4x4 supercell ($\delta$=0.015625) with a 3x3x3 Monkhorst-Pack kpoint mesh. Fig. S2.1 shows parabolic curves, with a larger energy scale than those in Fig. 2(a,b) in the main text due to the larger supercell size. The trends for vacancy formation energy with strain in Fig. S2.2 are qualitatively similar to those in Fig. 4(a) in the main text, but also have a different energy scale. This is not unexpected, as the defect interactions in LMO can be quite long range due to electrostatic and strain coupling. This result demonstrates that the defect properties derived in this work are specific to the concentrations studied and cannot necessarily be generalized to other defect concentrations. Convergence in the larger supercell is more challenging, resulting in curves in Fig. S2.1 and Fig. S2.2 that do not look as quadratic as their counterparts in the smaller supercell.

| a) LMO *aab* | b) LMO *aba* |
|---|---|

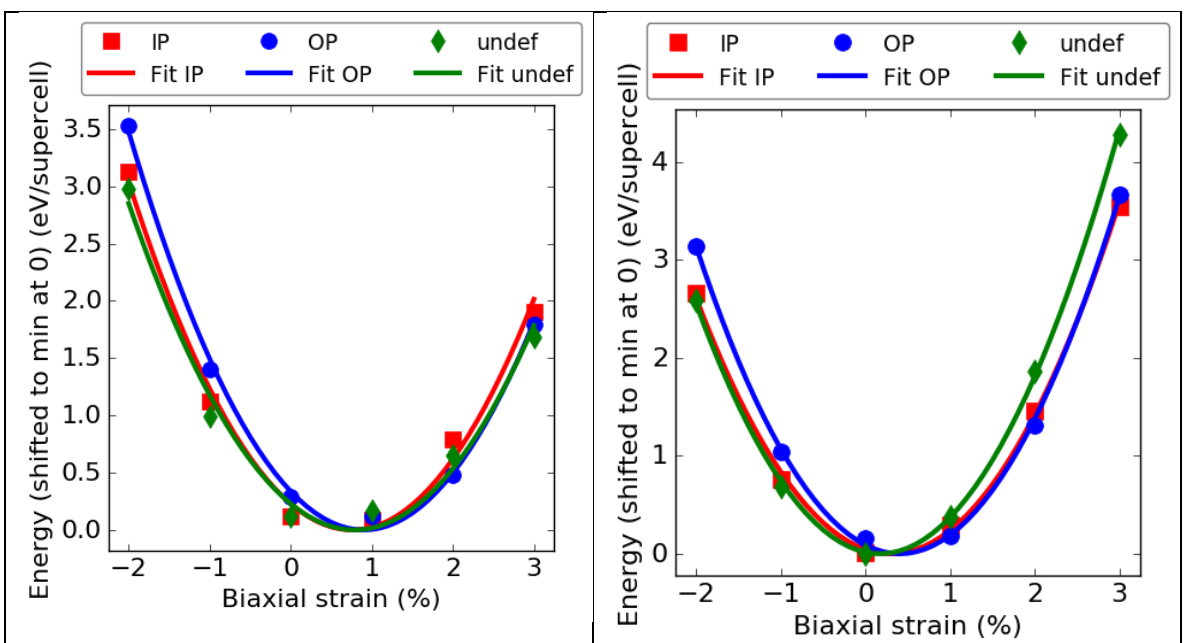

Fig. S2.1. Supercell energies for a 4x4x4 supercell, with energies adjusted to fit minimum is at 0 eV.

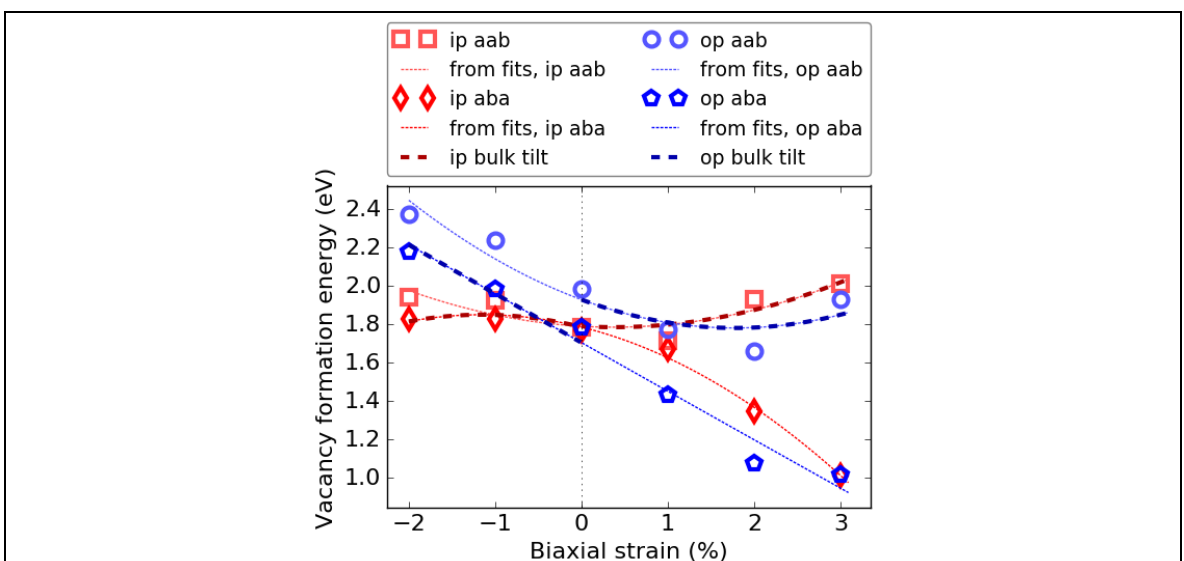

Fig. S2.2. LMO vacancy formation energy versus strain for a 4x4x4 supercell.

## S3. Magnetic structure

At lower temperatures, the B-site cations of the perovskites studied may adopt antiferromagnetic (AFM) rather than ferromagnetic (FM) structure.[2, 3] While this magnetic structuring may make a difference in trends and values for vacancy formation energy with strain, the overall model and conclusions in the main paper appear to continue to hold.

Fig. S3.1 shows raw energy curves for LaFeO$_3$ aab with a a G-AFM structure for its B-site cations. Fig. S3.1 shows that the G-AFM structure still produces parabolic curves in undefected and defected supercell energy with strain.

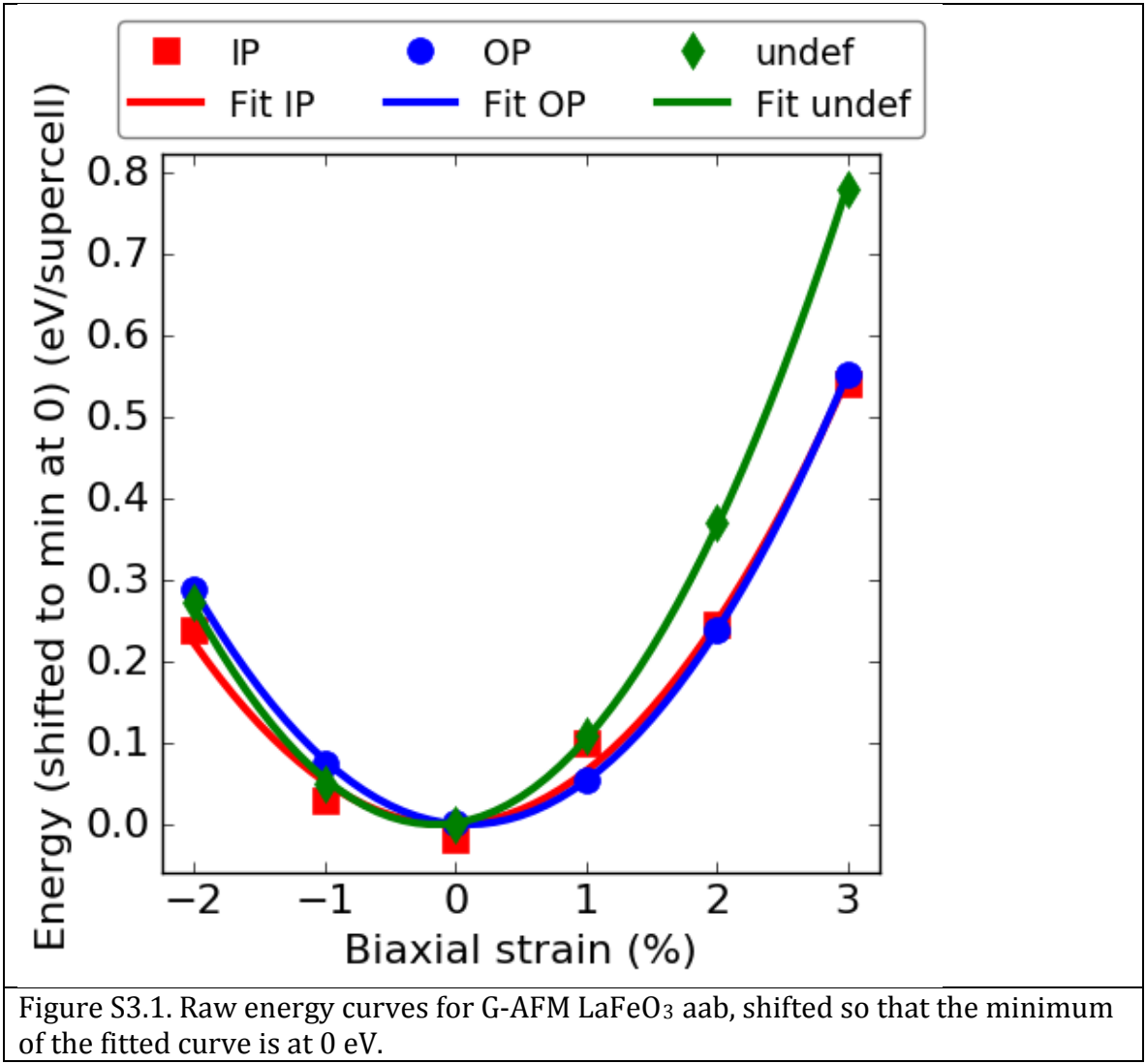

Figure S3.1. Raw energy curves for G-AFM LaFeO$_3$ aab, shifted so that the minimum of the fitted curve is at 0 eV.

Fig. S3.2 shows trends in vacancy formation energy with strain for the G-AFM and FM structures. Values in Fig. S3.2 are shifted to be zero at zero strain to compare the trends more clearly; the value of this shift (the zero-strain vacancy formation energy) is approximately 3.3 eV for the IP and OP zero-strain vacancy formation energy in the FM structure, and 3.8 eV in the G-AFM structure. Fig. S3.2 shows similar qualities for the trends in FM and G-AFM vacancy formation energies with strain. For each magnetic structure, the IP and OP curves show similar scales to each other. For both magnetic structures, the IP curve decreases more in compression than the OP curve, and all curves are concave down. The starting cubic volume was not re-relaxed for the G-AFM structure, so the only difference between the two structures is internal relaxation stemming from changes in magnetic configuration.

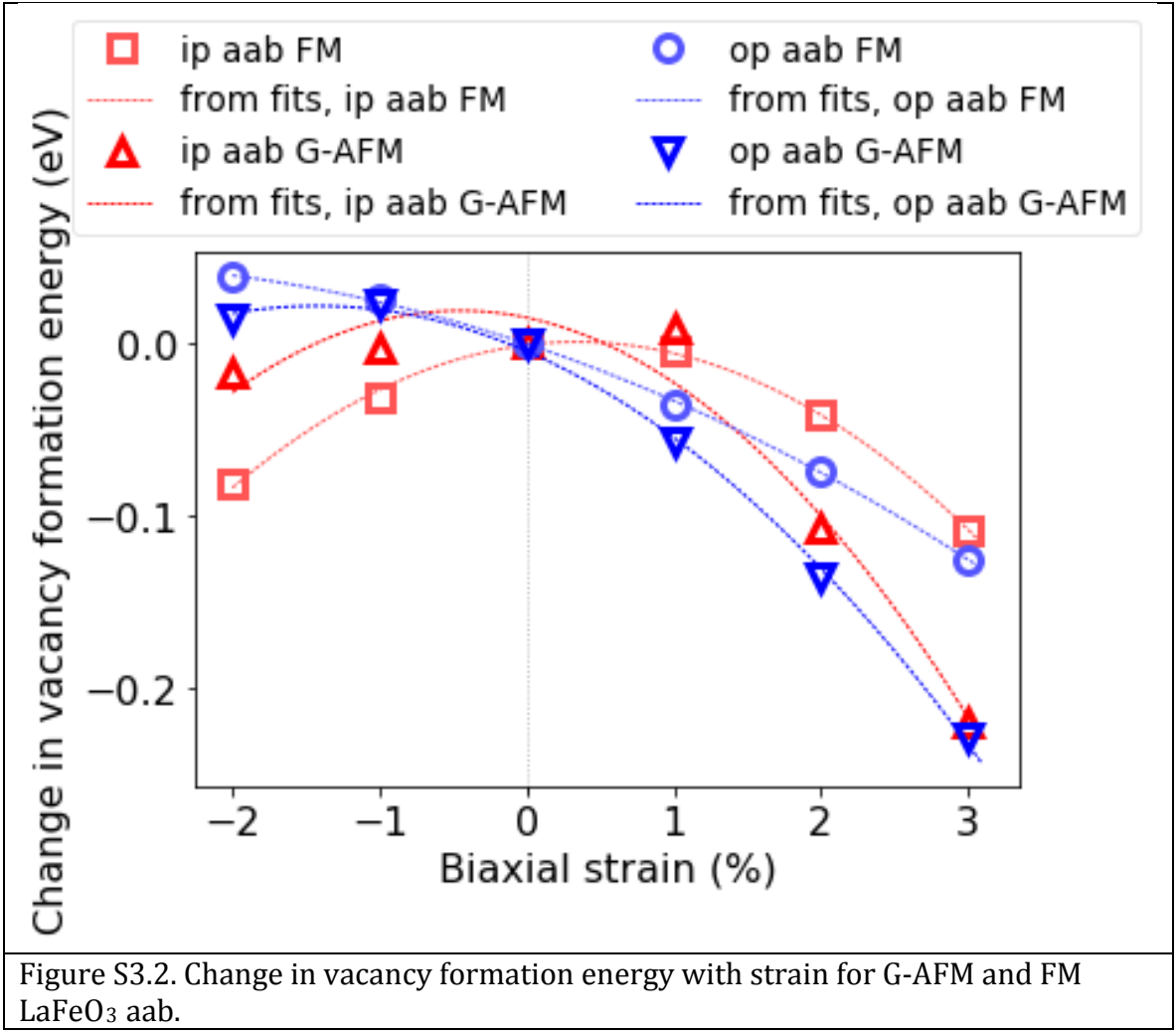

Figure S3.2. Change in vacancy formation energy with strain for G-AFM and FM LaFeO$_3$ aab.

## S4. Chemical expansion coefficient

Having collected vacancy relaxation volumes, we attempt to compare them to experiment via the chemical expansion coefficient. Chemical expansion is a quantity describing the volume change resulting from a chemical change in a material, for example, a change in non-stoichiometry for perovskites, as described, by, e.g., Bishop et al.[4] This quantity is distinct from the thermal expansion of a material, and combines with thermal expansion to produce total expansion.

Eq. S4.1 gives the chemical expansion coefficient defined by Bishop et al.[4] Eq. S4.2 derives the chemical expansion coefficient using the cubic vacancy relaxation volumes, where Δδ = 0.125. (Δδ from ABO$_3$ to ABO$_{3-δ}$ is equivalent to δ. For the 40-atom unit cell with a single oxygen vacancy, $(A,A')_8B_8O_{24-1} = (A,A')BO_{3-0.125}$, so $\delta = \Delta\delta = 0.125$.)

| | |
|---|---|
| $\alpha_C = \dfrac{\left.\Delta l/l_0\right|_T}{\Delta\delta}$ | Eq. S4.1 |
| $\alpha_C = \dfrac{\left.\left(V_{d,relaxed}^{1/3} - V_u^{1/3}\right)/V_u^{1/3}\right|_{T=0}}{\Delta\delta}$ | Eq. S4.2 |

Fig. S4 compares the experimental coefficients of chemical expansion, $\alpha_C$, with the coefficients derived from cubic vacancy relaxation volumes. The predicted values are much larger than experiment for LMO and overlap with the experimental range for LFO, although that range is quite large. In general the comparison is difficult due to limited data.

Temperature effects may be one source of error in the comparison. However, since chemical expansion coefficient decreases with decreasing temperature, about 0.003 per 100°C for LSGN between 700°C and 900°C,[5] and since the vacancy relaxation volume calculations are effectively performed at T=0, this error would be expected to lead to underestimation of chemical expansion coefficients from the calculations, not the observed overestimation. Thus temperature effects may be masking an even larger error from the calculations than is suggested by Fig. S4.

The higher value for LMO may be due to our larger vacancy concentration versus experiment, with $\delta=0.125$ versus $\delta<0.04$ in Ref. [6], which may affect our relaxation volumes. There may also be low spin versus high spin issues (all Mn are high spin in the calculations, but were identified as low spin in Ref. [6]), other temperature effects, or magnetic ordering effects. Further comparison for more systems is needed to assess the accuracy of the models and potential sources of error.

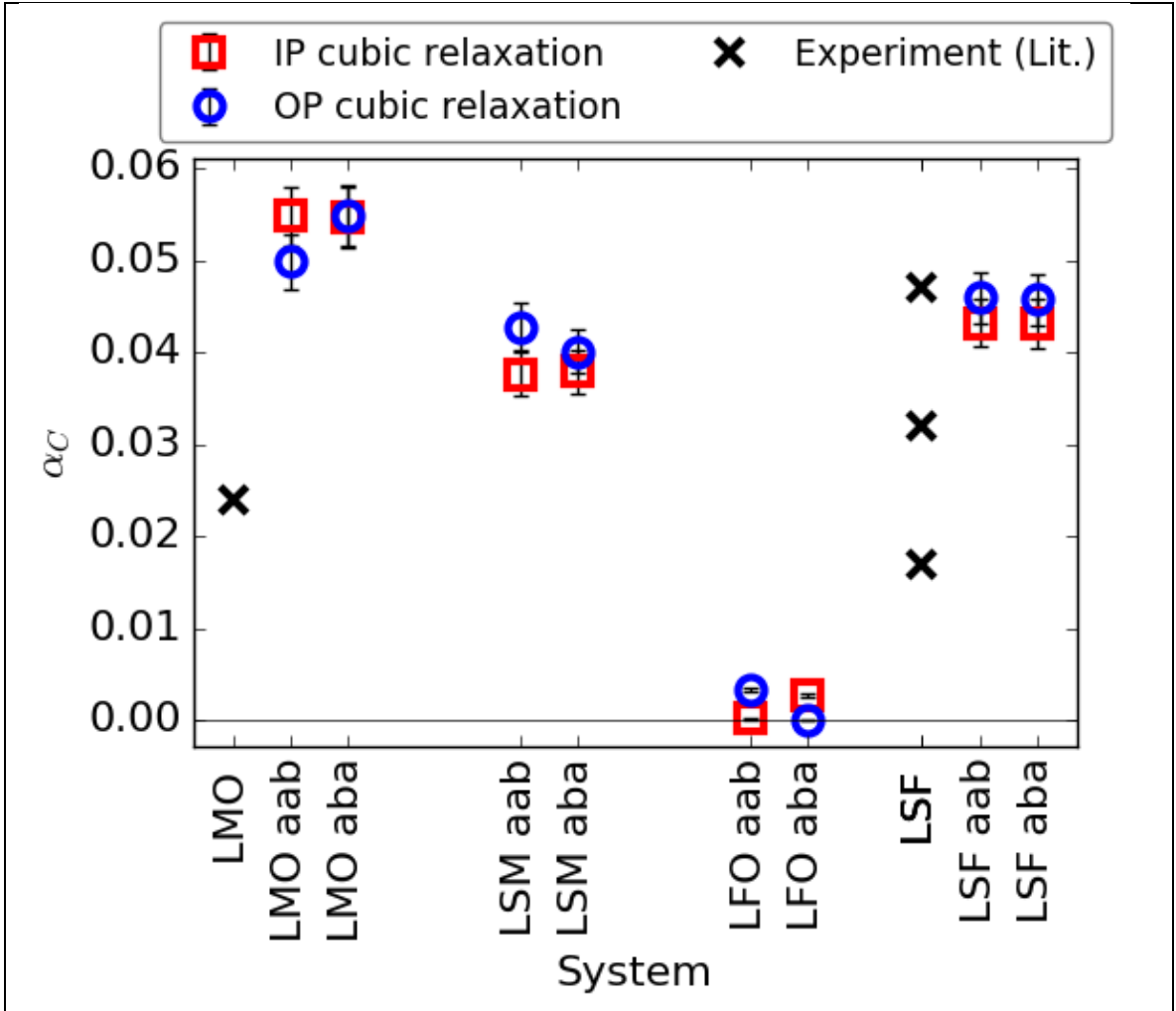

Fig. S4. Chemical expansion coefficient[4] $\alpha_C$ compared with analogous values derived from the cubic vacancy relaxation volume. Experimental values are taken from Ref. [4], where LMO is $LaMnO_{3-\delta}$ and LSF is actually $La_{0.25}Sr_{0.75}FeO_{3-\delta}$, where the different ratio of A-site cations may not necessarily provide a good comparison to $La_{0.75}Sr_{0.25}FeO_{3-\delta}$. Error bars on experiment are not given but are likely very small, near the +/- 0.002 in the related Ref. [5] and with generally high-quality fits, e.g. in Ref. [6].

If a strain-constrained $\alpha_C$ were instead derived from the vacancy formation volumes of Fig. 8 in the main text, the IP $\alpha_C$ would be similar to the calculated values in Fig. S4, while the OP $\alpha_C$ would be even larger. Since Fig. 4 in the main text shows lower vacancy formation energies for OP vacancies in most systems at 2% and 3% tensile strain, OP vacancies may be slightly more dominant in the strained material than IP vacancies, for most systems. Therefore, if dominated by OP vacancies, an analogous $\alpha_C$ value for a strained material may be larger than its unconstrained counterpart.

## S5. Transforming supercells

This method transforms an orthorhombic or trigonal supercell with lattice vectors $a$, $b$, and $c$, into a 2x2x2 (40-atom) semi-cubic supercell, e.g. with lattice vectors $a'$, $b'$, and $c'$ close to axes $x$, $y$, and $z$. The general method is outlined briefly in Ref. [7] and uses:

| | |
|---|---|
| $U' = P(U - T)$ | Eq. S5.1 |
| $P = M^{-1}$ | Eq. S5.2 |

where $U$ is a column vector of fractional coordinates in the untransformed system, $U'$ is a column vector of fractional coordinates in the transformed system, $M$ is a matrix transforming the old lattice parameters into the new lattice parameters (see below), and $T$ is a translation vector describing the origin of the new lattice parameters in terms of fractional coordinates in the untransformed lattice parameters.

1. Identify matrix M such that:

| | |
|---|---|
| $\begin{bmatrix} a_x & b_x & c_x \\ a_y & b_y & c_y \\ a_z & b_z & c_z \end{bmatrix} M = \begin{bmatrix} a'_x & b'_x & c'_x \\ a'_y & b'_y & c'_y \\ a'_z & b'_z & c'_z \end{bmatrix}$ | Eq. S5.3 |
| $LM = L'$ | Eq. S5.4 |

If matrix $M$ is difficult to identify, then do the following:
1a. Identify two atoms, in this example two A-site La atoms, which will form lattice vector $a'$ in the semi-cubic system. One of these La atoms will be at the origin of the new coordinate system, which we will label for future use as $La_0$. Identify their (x,y,z) coordinates using a combination of VESTA visualization[1] and an XYZ file (.xyz) generated by a VESTA Data Export. Subtract their coordinates to form lattice vector $a'$, doubling the vector if necessary if those La were neighboring La rather than one La apart.
1b. Calculate the first column of $M$ as follows:

| | |
|---|---|
| $\begin{bmatrix} a_x & b_x & c_x \\ a_y & b_y & c_y \\ a_z & b_z & c_z \end{bmatrix} \begin{bmatrix} A \\ B \\ C \end{bmatrix} = \begin{bmatrix} a'_x \\ a'_y \\ a'_z \end{bmatrix}$ | Eq. S5.5 |
| $L \begin{bmatrix} A \\ B \\ C \end{bmatrix} = \begin{bmatrix} a'_x \\ a'_y \\ a'_z \end{bmatrix}$ | Eq. S5.6 |
| $\begin{bmatrix} A \\ B \\ C \end{bmatrix} = L^{-1} \begin{bmatrix} a'_x \\ a'_y \\ a'_z \end{bmatrix}$ | Eq. S5.7 |

1c. Repeat the process for the other columns of $M$.
2. Convert $M$ to integer values for use in pymatgen. Use lists as column vectors for the pymatgen supercell transformation method:

| | |
|---|---|
| $M_{Pbnm,\#62} = \begin{bmatrix} 1 & 1 & 0 \\ -1 & 1 & 0 \\ 0 & 0 & 1 \end{bmatrix}$ | Eq. S5.8 |
| $M_{R\bar{3}c,\#167} = \begin{bmatrix} 1.33 & 0.66 & 0.66 \\ 0.66 & 1.33 & -0.66 \\ -0.33 & 0.33 & 0.33 \end{bmatrix} \rightarrow \begin{bmatrix} 4 & 2 & 2 \\ 2 & 4 & -2 \\ -1 & 1 & 1 \end{bmatrix}$ | Eq. S5.9 |
| $M_{a^-a^-b^+ to\ a^-b^+a^-} = \begin{bmatrix} 1 & 0 & 0 \\ 0 & 0 & -1 \\ 0 & 1 & 0 \end{bmatrix}$ | Eq. S5.10 |
| For Pbnm, SupercellTransformation[[1,1,0],[-1,1,0],[0,0,1]] | Eq. S5.11 |
| For R-3c, SupercellTransformation[[4,2-1],[2,4,1],[2,-2,1]] | Eq. S5.12 |

3. Visualize the new supercell in VESTA.
4. Use a pymatgen translation so that $La_0$ will be placed close to the origin in the new system, e.g. SymmOp.from_rotation_and_translation with
translation_vec=[0.25, 0.25, 0.25] for R-3c or
translation_vec=[0, 0, 0.25] for Pbnm
No translation necessary for switch of axes from $a^-a^-b^+$ to $a^-b^+a^-$ tilt system.

Optional, 5. Double check an atom:
5a. Identify *T* as the fractional coordinates of $La_0$ in the untransformed system, e.g. (0, 0, 0.25) for Pbnm or (0, 0, 0.25) for R-3c.
5b. Pick an atom whose coordinates are approximately known in the desired new system, e.g. a B-site cation at fractional coordinates (0.25, 0.25, 0.25), as a column vector *U'*.
5c. Use a combination of VESTA visualization and an XYZ file (as above) to identify the fractional coordinates of that exact B-site cation in the untransformed system, e.g. (0.66, 0.33, 0.33) for R-3c using the choices leading to the transformations above, as a column vector *U*.
5d. Calculate *P* as the inverse of *M*.
5e. Verify that Eq. S5.1 is true.

6. Adjust the axes so that they lie approximately along *x*, *y*, and *z*. If this adjustment is not obvious, do the following:
6a. Have pymatgen print out the structure as a POSCAR-type file. The lattice parameter section of the POSCAR-type file will show the *x*, *y*, and *z* values of the transformed lattice parameters.
6b. Get the magnitude of each lattice vector:

| | |
|---|---|
| $\left\|\vec{a'}\right\| = \sqrt{a'^2_x + a'^2_y + a'^2_z}$ | Eq. S5.13 |

These magnitudes should all be approximately equal. Form a diagonal matrix *D* using these magnitudes. Form the transformed lattice parameter matrix *L'* using the transformed lattice parameters as column vectors (note that they are row vectors in the POSCAR file). Calculate the rotation matrix as follows:

| $$D = \begin{bmatrix} \lvert \vec{a'} \rvert & 0 & 0 \\ 0 & \lvert \vec{b'} \rvert & 0 \\ 0 & 0 & \lvert \vec{c'} \rvert \end{bmatrix}$$ | Eq. S5.14 |
|---|---|
| $D = QL'$ | Eq. S5.15 |
| $DL'^{-1} = Q$ | Eq. S5.16 |

6c. Enter $Q$ as a rotation matrix for a pymatgen; enter it as shown so that the first row is the first list. For example, for R-3c, use:
bigmat=[[0.7112, 0.4106, -0.5708],[0, 0.8213, 0.5708],[0.7112, -0.4106, 0.5708]]
my_rot_op = SymmOp.from_rotation_and_translation(rotation_matrix=bigmat)
For Pbnm, use:
  bigmat=[[0.719519820006603, 0.695314786989596, 0],
      [ -0.719519820006603, 0.695314786989596, 0],
      [0,0,1]]
For switching tilt system order, use:
bigmat=[[1,0,0],[0,0,1],[0,-1,0]]
6d. Have pymatgen save off the POSCAR file.

7. Adjust the number of atoms. Visualize the supercell in VESTA and make sure it looks correct. Use VESTA's Select tool and the Backspace or Delete key to remove all but the appropriate 40 atoms (8 A-site, 8 B-site, and 24 oxygen atoms).
7a. Export the VESTA data as an XYZ file. Do not export hidden atoms. (This option is not available for exporting directly to a POSCAR file.)
7b. Copy the XYZ file. Manually change it into a POSCAR-type file, using the correct lattice parameters, with the correct scaling (e.g. if the original lattice parameters were three times as large, scale down by 3.)
7c. Open the POSCAR file in VESTA. Export Data as a POSCAR file with fractional coordinates.

## S6. Addressing strain with perovskites

This section discusses approaches to biaxial strain in the computational literature and will help guide future researchers in understanding what choices have made in different contexts.

Table S6.1. Strain approaches
| Reference | System | Strain | Defects | Study |
|---|---|---|---|---|
| Zhang[8] | STO, tetragonal at low T, a-00? | hydrostatic pressure from PSTRESS | fixed to undefected lattice parameters | Evf |
| Yang[9] | BTO, tetragonal | cell shape kept tetragonal for | fixed to undefected | Emig |

| | | external strain; biaxial strain along a and b with relaxation along c; (strain x and y, relax z) | lattice parameters | |
|---|---|---|---|---|
| Tealdi[10] | LGO, orthorhombic Pbnm | ab strain, c relax (orthorhombic) | introduced into optimized structure | Emig, (GULP for structures) MD |
| Tahini[11] | SCO, ideally Pm-3m, orthorhombic in practice cubic for defect-free | Specifics not mentioned! | no comment; probably no re-relaxation | Emig, Evf |
| Petrie[12] | SCO, orthorhombic | Specifics not mentioned | | |
| May[13] | LNO, R-3c, a-a-c-; first-order transition observed at 0.5% strain | imposed in-plane lattice parameter equality to mimic a substrate, relax in c | None | Octahedral bond lengths and angles |
| Lee[14] | LMO, R-3c, Pbnm | square-lattice substrate, c-lattice parameter relaxation; epitaxially-constrained Pbnm (c or ab); note that the square lattice substrate does not fall along a and b (see Gunter 2012[15] for a better diagram) | | Phase sequence under strain; FM more square base than AFM; alternative low-energy low-orthorhomb state |
| Jalili[16] | LSM | cubic lattice constant used, 001 (La,Sr)O terminated surace; strain in x and y, relax in z | | Eseg, Evf |
| Han[17] | LSC 214 (as | [100] and [010] | | Evf, Emig, |

| | tetragonal) and 113 (as cubic) | strain, [001] relaxation | | Eincorp |
|---|---|---|---|---|
| Guo[18] | LSM, which can be treated as tetragonal | fixed IP lattice params (to SXRD and RSM values) | c-lattice param recalculated | Electronic structures |
| Guunter[15] | CMO, very orthorhombic Pnma | cubic imposition fixed to substrate in two directions, relaxation in third direction; angle between a and c when strain is along b is allowed to relax | | transverse-optical modes |
| Gan[19] | LMO, LSM | cubic lattice constant; strain along x and y, relax in z | | Evf, Emig Note that position of Ovac matters, and they are not parabolic |
| Eklund[20] | CTO, Pbnm; HT cubic | as with Ref. [15]; looked at tilting of t_c but dismissed it | | energy, ferroelectricity |
| Zayak[21] | SrRuO3, orthorhombic | depending on orientation, an orthorhombic epitaxial film may or may not develop a tilt angle with respect to the substrate (see Fig. 5) | | |
| Rondinelli 2009[22] | LCO, rhombohedral | fix in-plane pseudocubic lattice paramter; allow c to relax; uniaxial tension also applied in c; rhombohedral angle allowed to adjust, but I think the pseudocubic | | Structural effects |

| | | angle is then 90 degrees | | |
|---|---|---|---|---|
| Rondinelli 2010[23] | SFO, ideal cubic | a, b fixed; c relax | | Structural effects, electronic effects |
| Donner[24] | LCO, LSC | not mentioned; probably cubic/tetragonal | | Evf |
| Aschauer[25] | CMO, Pnma | Constrain in plane of equal a and c axes, relax b-axis length and internal coordinates; specification of length implies fixed angle? | | |
| Akhade 2011[26] | LaBO3, SrBO3 | All cubic, 1x1x1 cells | | bulk electronic structure |
| Akhade 2012[27] | LaBO3, SrBO3 | 1x1 footprint; I think these are cubic, from previous paper; not explicitly mentioned | | surface electronic structure and Evf |
| Han 2011[28] | LCO | cubic lattice constants | | Evf, Emig |

Experimental papers[29, 30]

Table S6.2. Tallies for strain approaches

| Method | Number |
|---|---|
| Specifically mention using a cubic or tetragonal cell | From experiment or matching to experimental lattice: [8, 16-19, 23, 28] From convenience:[26] [9] |
| Specifically mention fixing a square footprint for an orthorhombic or rhombohedral cell | [13-15, 20-22, 25] |
| Does not fix a square footprint for an orthorhombic or other cell (a<>b) | [10] |
| Does not mention cell specifics | [11, 12, 24, 27] |
| Specifically mentions an out-of-plane angle with respect to the plane of strain | [15, 20, 21] |

A survey of 21 studies shows several main ways of addressing strain with perovskites. Nine of these studies use cubic or tetragonal cells, either for simplicity (2 studies)[9, 26] or by citing a cubic or tetragonal experimental lattice parameter (7 studies).[8, 16-19, 23, 28] Eight studies use an orthorhombic or rhombohedral cell, most specifically fixing a square footprint, at least in one orientation with respect to the substrate (7 studies),[13-15, 20-22, 25] and only one study allowing $a \neq b \neq c$.[10] Of the orthorhombic cell studies, three studies make specific mention of allowing a non-orthogonal out-of-plane angle. [15, 20, 21] Four studies could not be definitively categorized. [11, 12, 24, 27] In summary, only 3 out of 21 studies felt the need to specifically mention an out-of-plane angle: Eklund, Fennie, and Rabe state that the angle did not lower energy for a large-strain test case, and subsequently did not use a non-orthogonal angle;[20] Günter *et al.* mention an out-of-plane angle only for a uniaxial strain case;[15] and Zayak *et al.* mention an out-of-plane angle only for epitaxial layers oriented along the [110] direction for a *Pbnm* unit cell (4 formula units).[21]